\documentclass[sigconf,natbib=True]{acmart}

\usepackage{booktabs} 
\usepackage{algorithm}
\usepackage{algorithmic} 

\usepackage{subfigure}
\usepackage{graphics}
\usepackage{ifthen}
\usepackage{amsmath} 
\usepackage{latexsym}
\usepackage{CJK}
\usepackage{amsfonts}
\usepackage{ifthen}
\usepackage{caption}
\usepackage{graphicx}
\usepackage{float}
\usepackage{multirow}
\usepackage{subfigure}
\usepackage{balance}
\usepackage{subfigure}
\usepackage{graphics}
\usepackage{enumerate}
\usepackage{verbatim}
\usepackage{graphicx}
\usepackage{float}
\usepackage{booktabs}
\usepackage[hang,flushmargin]{footmisc}
\usepackage{caption}
\usepackage{bbold}
\usepackage{color}
\usepackage{xspace}

\newcommand{\paratitle}[1]{\vspace{1.5ex}\noindent\textbf{#1}}

\newcommand{\ignore}[1]{}

\newcommand{\baby}{\textsc{BBQRec}\xspace}
\AtBeginDocument{%
  }

\setcopyright{acmlicensed}
\copyrightyear{2018}
\acmYear{2018}
\acmDOI{XXXXXXX.XXXXXXX}
\acmConference[Conference acronym 'XX]{Make sure to enter the correct
  conference title from your rights confirmation emai}{June 03--05,
  2018}{Woodstock, NY}
\acmISBN{978-1-4503-XXXX-X/18/06}

\begin{document}

\title{BBQRec: Behavior-Bind Quantization for Multi-Modal Sequential Recommendation}

\author{Kaiyuan Li}
\affiliation{%
  \institution{Kuaishou Technology}
  \city{Beijing}
  \country{China}}
\email{likaiyuan03@kuaishou.com}

\author{Rui Xiang}
\affiliation{%
  \institution{Kuaishou Technology}
  \city{Beijing}
  \country{China}}
\email{xiangrui@kuaishou.com}

\author{Yong Bai}
\affiliation{%
  \institution{Kuaishou Technology}
  \city{Beijing}
  \country{China}}
\email{baiyong@kuaishou.com}

\author{Yongxiang Tang}
\affiliation{%
  \institution{Kuaishou Technology}
  \city{Beijing}
  \country{China}}
\email{tangyongxiang@kuaishou.com}

\author{Yanhua Cheng}
\affiliation{%
  \institution{Kuaishou Technology}
  \city{Beijing}
  \country{China}}
\email{chengyanhua@kuaishou.com}

\author{Xialong Liu}
\affiliation{%
  \institution{Kuaishou Technology}
  \city{Beijing}
  \country{China}}
\email{zhaolei16@kuaishou.com}

\author{Peng Jiang}
\affiliation{%
  \institution{Kuaishou Technology}
  \city{Beijing}
  \country{China}}
\email{jiangpeng@kuaishou.com}

\author{Kun Gai}
\affiliation{%
  \institution{Kuaishou Technology}
  \city{Beijing}
  \country{China}}
\email{gai.kun@qq.com}
\renewcommand{\shortauthors}{Kaiyuan Li et al.}

\begin{abstract}
Multi-modal sequential recommendation systems leverage auxiliary signals (e.g., text, images) to alleviate data sparsity in user-item interactions. While recent methods exploit large language models to encode modalities into discrete semantic IDs for autoregressive prediction, we identify two critical limitations: (1) Existing approaches adopt fragmented quantization, where modalities are independently mapped to semantic spaces misaligned with behavioral objectives, and (2) Over-reliance on semantic IDs disrupts inter-modal semantic coherence, thereby weakening the expressive power of multi-modal representations for modeling diverse user preferences.  

To address these challenges, we propose a \textbf{B}ehavior-\textbf{B}ind multi-modal \textbf{Q}uantization for Sequential \textbf{Rec}ommendation (BBQRec for short) featuring \textbf{dual-aligned quantization} and \textbf{semantics-aware sequence modeling}. First, our behavior-semantic alignment module disentangles modality-agnostic behavioral patterns from noisy modality-specific features through contrastive codebook learning, ensuring semantic IDs are inherently tied to recommendation tasks. Second, we design a discretized similarity reweighting mechanism that dynamically adjusts self-attention scores using quantized semantic relationships, preserving multi-modal synergies while avoiding invasive modifications to the sequence modeling architecture. Extensive evaluations across four real-world benchmarks demonstrate \baby's superiority over the state-of-the-art baselines. 
\end{abstract}




\maketitle

\section{Introduction}
Sequential recommendation, which aims to model evolving user preferences through behavior sequences, has become a cornerstone of modern recommender systems~\cite{HuangZDWC18, ugrec, dev_1}. Although existing studies~\cite{dev_2, iim} demonstrate that the capture of sequential dependencies significantly improves intention prediction, their effectiveness is heavily based on sufficient interaction data, an assumption frequently violated in real-world scenarios. The inherent sparsity of user behavior data forces current approaches to depend on coarse-grained sequential patterns creating a fundamental bottleneck that limits recommendation performance.
\begin{figure}[htbp]
\centering
\includegraphics[scale=0.49]{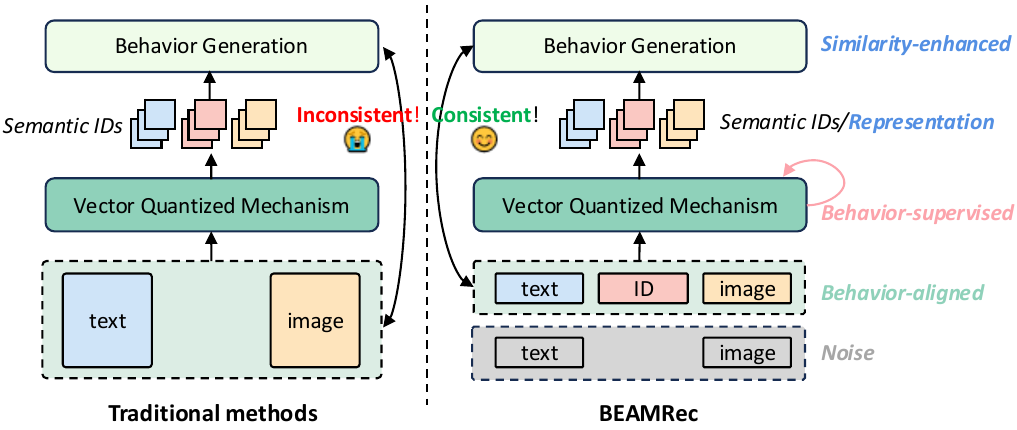}
\caption{A simple diagram illustrates the key differences between our proposed method and existing approaches.}
\label{fig:intro}
\end{figure}

To alleviate the sparsity problem of recommendation, many research efforts have been devoted to exploiting context information such as textual descriptions\cite{rns}, visual images\cite{vbpr}, etc. A common paradigm of these solutions is to transform these heterogeneous information into generic feature vectors, which are fed into a supervised learning model\cite{fdsa,unisrec} together with sequential representations to predict the user's next interest.

Inspired by the success of generative models in the field of large language models (LLMs), generative recommendation has been a common strategy for sequential recommendation\cite{tiger,letter,cost,eager}. Unlike traditional matching methods, this paradigm employs end-to-end generative models to directly predict next-item's semantic IDs in an autoregressive manner. Specifically, this approach consists of two stages: item tokenization and autoregressive generation. The purpose of item tokenization is to learn the discrete semantic IDs of items. For the autoregressive generation task, the encoder-decoder architecture is the most widely used backbone due to its excellent capabilities in sequence modeling and generation~\cite{tiger,letter,cost}.

Despite achieving impressive results, these methods often rely on quantifying items using only a single type of content information during the item tokenization stage\cite{tiger,letter,cost}, lacking the application of multi-modal information. Meanwhile, most of these methods overlook the inconsistency of these content information in behavior generation tasks. Two items that are similar in images or texts may have completely unrelated behavioral pattern (such as mobile phones from various brands). This misalignment limits further advances in generative recommendation. To address this issue, recent works attempt to introduce collaborative signals\cite{letter} and multi-modal information\cite{eager,mmgrec} aiming to alleviate these problems. 

Although effective, we argue that there remain several critical challenges for multi-modal utilization in generative sequential recommendation: \textbf{(1) Behavior-Semantic Alignment:} Current research\cite{eager,mmgrec} overlooks quantization strategy for effectively aligning multi-modal information with behavioral signals. Prior studies \cite{xv2024improving,mvgcn, maris,multimodal-suvery} reveal that modalities often contain behavior-irrelevant noise. Therefore, disentangling behavior-irrelevant information is crucial but few works in generative sequential recommendation address this challenge; \textbf{(2) Non-Invasive Sequence Modeling:} Most existing approaches\cite{tiger,cost,eager,mmgrec} retain only semantic IDs while discarding semantic representations in the generation stage, leading to the loss of semantic signals learned from pre-trained representations. However, directly integrating pre-trained representations in attention mechanisms has been repeatedly shown to limit recommendation performance \cite{nova,difsr,asif}. A suitable non-invasive methodology for incorporating such information remains unresolved in generative sequential recommendation.

To tackle these challenges, we propose a novel \textbf{B}ehavior-\textbf{B}ind multi-modal \textbf{Q}uantization for Sequential \textbf{Rec}ommendation (\baby)  and Figure \ref{fig:intro}
shows the key difference between \baby and traditional methods. Firstly, to enhance the alignment between the semantic IDs and the behavioral generation task, a mutual information minimization module is designed to decouple behavior-aligned signals from modality-specific features. These aligned features are then encoded through a shared behavior-dominated codebook, and subsequently optimized via sequence-item contrastive learning while jointly reconstructing with modality-specific representations.
Secondly, to preserve behavior-aligned signals, we develop a similarity-enhanced self-attention mechanism that avoids discarding or directly inheriting quantized representations by discretizing pairwise item similarities from learned embeddings, then dynamically reweighting standard self-attention scores through these semantic relationships. This design enhances sequence modeling capabilities in a non-invasive manner while maintaining the original attention architecture.
We construct extensive experiments on four datasets and experimental results show that our proposed \baby can significantly outperform all the baselines. To summarize, the contributions of this paper are listed as follows.
\begin{itemize}
\item We propose a novel framework for multi-modal quantization and focus on the two key challenges of generative sequential models: behavior-semantic alignment and non-invasive sequence modeling.

\item To align the multi-modal information with behavior information, we propose a quantization strategy that decouples behavior-aligned features from modality-specific noise.

\item To preserve the behavior-aligned information learned from item tokenization stage, we propose a novel auxiliary module within the self-attention through a non-invasive manner.

\item Extensive experiments on four real-world datasets demonstrate that our model obtains a significant performance improvement over the state-of-the-art sequential recommendation models.
\end{itemize}

\section{Related Work}
In this section, we provide a brief overview of the related work from two perspectives, including sequential recommendation, multi-modal recommendation, respectively.

\paratitle{Sequential Recommendation}. 
The goal of sequential recommendation is to identify meaningful patterns within sequences efficiently. Early research in this area mainly employed Markov chain models~\cite{fpmc, hrm} to capture lower-order sequential dependencies. However, with the rise of deep learning, Recurrent Neural Networks~\cite{gru, narm, stamp} and self-attention models~\cite{sasrec, bert4rec, transformer4rec} have been utilized in various sequential modeling tasks to overcome the limitations of Markov models. More recently, pre-training methods~\cite{s3rec, unisrec, vqrec} have been explored to uncover intrinsic data correlations, thereby enhancing sequential recommendation performance.

Recently, influenced by generative AI, the generative sequential recommendation has started to gain more attention\cite{p5,vip5,tiger}. For example, TIGER\cite{tiger} encodes item attribute information into semantic IDs using Residual Quantized Variational Autoencoder (RQ-VAE)\cite{rqvae} and generates items in an autoregressive manner. Based on this work, LETTER\cite{letter} and EAGER\cite{eager} aims to introduce users' collaborative signals to assist the quantization process where CoST\cite{cost} focuses on how to leverage contrastive quantization to enhance semantic IDs. In this work, we focus on the tokenization strategy of semantic IDs in generative sequential recommendation.

\paratitle{Multi-modal Recommendation}. 
Utilizing multi-modal data, such as texts and images, has proven to enhance the effectiveness of recommendations across various
applications~\cite{modal_suvery_1, modal_suvery_2}. This approach has been particularly beneficial in areas like news recommendation \cite{miner,gnr}, micro-video recommendation\cite{micro_video_1,micro_video_2} and fashion recommendation\cite{fashion_rec_1,fashion_rec_2}. To fully exploit the interactions among these features, previous works usually used the matrix factorization technique~\cite{vbpr, taper} for multi-modal recommendation. Recently, deep learning techniques have also been utilized for multi-model recommendation. For example, MMGCN\cite{mmgcn} and GRCN\cite{grcn} build
modality graphs and use GCN\cite{gcn} to aggregate information.
MMGCL\cite{mmgcl} and MISSRec\cite{missrec} apply contrastive learning to discover the relationship between the user and the item. In this work, we focus on enhancing the effective utilization of multi-modal information in generative sequential recommendation.

\begin{figure*}[h!]
\centering
\includegraphics[scale=0.52,clip=true]{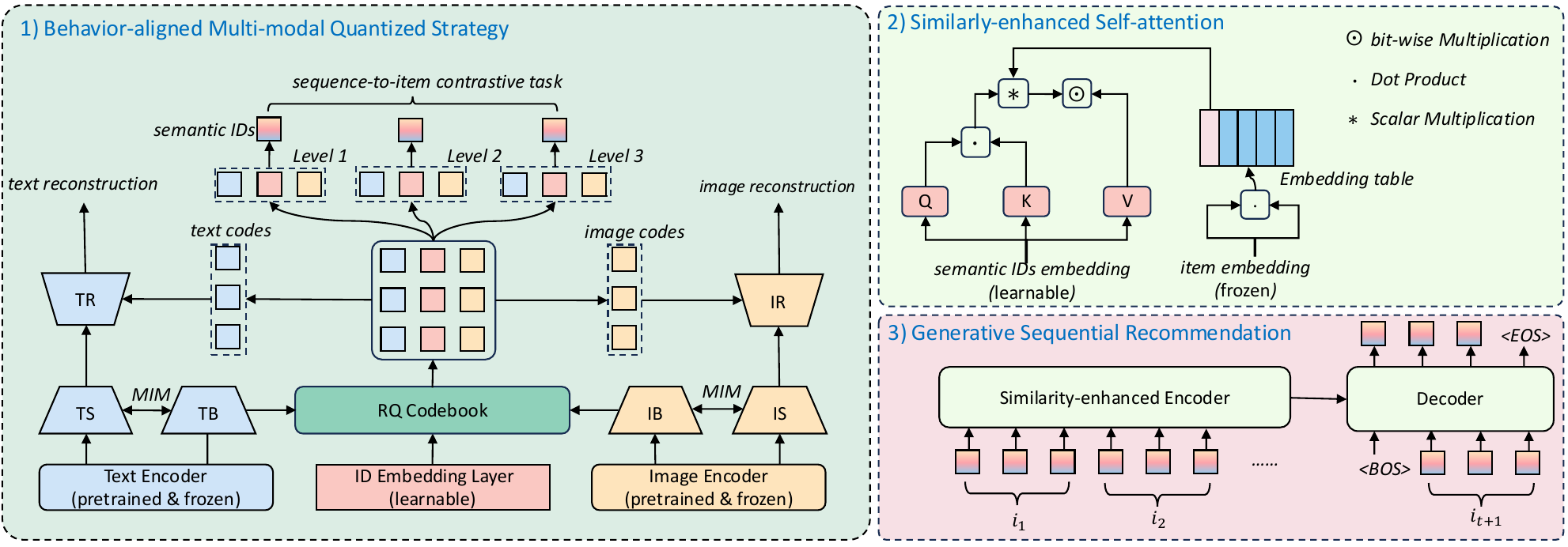}
\caption{The overall architecture of Behavior-aligned Multi-modal Quantized framework for Sequential Recommendation~(\baby for short). TS/IS refers to the text/image-specific encoder, TB/IB denotes the behavior-aligned text/image encoder, and TR/IR represents the text/image decoder.}
\label{fig:model}
\end{figure*}

\section{Preliminary}
\paratitle{Notations}.
Let $\mathcal{U}$ denote a set of users and $\mathcal{I}$ denote a set of items, where $|\mathcal{U}|$ and $|\mathcal{I}|$ are the numbers of users or items.
For each user $u\in \mathcal{U}$, we use
$S^u_{t}=\{i^u_1, i^u_2, \cdots, i^u_t \}$ to represent the sequence of interaction of the elements, where $i^u_k$ represents the element with which $u$ has interacted at the $k$-th time step. For simplicity, we
describe the approach for a single user and drop the superscript $u$ for ease of reading.

\paratitle{Task Definition}. In the item generation task, each item $i$ is encoded as a discrete semantic ID sequence with length of $L$: $i = [c_1, c_2, \cdots, c_L]$. Each $c_j$ represents the level of the semantic IDs. When employing RQ-VAE for semantic ID sequence generation, the hierarchical structure exhibits a coarse-to-fine progression across quantization levels from 
$c_1$ to $c_L$, where lower-index codes capture broad semantic patterns while higher-index codes encode granular behavioral details. Given a user interaction sequence $S_t = [i_1, i_2, \cdots, i_t]$, the aim is to predict the next item $i_{t+1}$ by modeling $P(i_{t+1} \mid S_t)$. We use an autoregressive model to predict the sequence of semantic IDs for $i_{t+1}$:
\begin{equation}
P(i_{t+1} \mid S_t) = \prod_{j=1}^{L} P(c_j \mid c_1, \cdots, c_{j-1}, S_t)
\end{equation}
This formulation captures hierarchical item dependencies through the conditional probability $P(c_j \mid c_1, \cdots, c_{j-1}, S_t)$ of generating the $j$-th semantic ID given preceding IDs and session context $S_t$, achieving accurate predictions via coarse-to-fine pattern.

\section{Methodology}
In this section, we introduce the proposed \textbf{BE}havior-\textbf{A}ligned \textbf{M}ulti-modal Quantized method for Sequential \textbf{Rec}ommendation (\baby) in detail, and the overall architecture of \baby is presented in Figure ~\ref{fig:model}.
In the following, we start with a behavior-aligned module, after this, we present the similarity-enhanced self-attention.

\subsection{Behavior-aligned Multi-modal Quantized Strategy}
In this section, we will introduce our proposed behavior-aligned module. Specifically, we will present it in four parts. First, we introduce our behavior disentanglement module, followed by the behavior quantization module, and then a sequence-to-item behavior contrastive learning module. Finally, we introduce the total loss function of these module.

\subsubsection{Behavior Disentanglement Module}
The behavior disentanglement module is designed to capture semantic information aligned with the behavior for each modality. To achieve this goal, we first deploy two encoders: a modality-specific encoder and a behavior-aligned encoder: Let $\mathbf{x}_i^{\text{text}}$ and $\mathbf{x}_i^{\text{image}}$ represent the inputs from text and image modalities of item $i$, respectively. The modality-specific encoders, $\mathbf{Encoder_{TS}}$ and $\mathbf{Encoder_{IS}}$, map the pre-trained modality embedding into modality-specific representations $\overline{\mathbf{z}}_i^{\text{text}}$, $\overline{\mathbf{z}}_i^{\text{image}}$ which is not relevant to the behavior information. The behavior alignment encoders, $\mathbf{Encoder_{TB}}$ and $\mathbf{Encoder_{IB}}$, encode these inputs into a behavior-aligned latent space, resulting in representations $\mathbf{z}_i^{\text{text}}$ and $\mathbf{z}_i^{\text{image}}$:
\begin{equation}
\begin{split}
\overline{\mathbf{z}}_i^{\text{text}} & = \mathbf{Encoder_{TS}}(\mathbf{x}_i^{\text{text}}) ; 
\mathbf{z}_i^{\text{text}}  = \mathbf{Encoder_{TB}}(\mathbf{x}_i^{\text{text}}) \\
\overline{\mathbf{z}}_i^{\text{image}} & = \mathbf{Encoder_{IS}}(\mathbf{x}_i^{\text{image}}) ; 
\mathbf{z}_i^{\text{image}}  = \mathbf{Encoder_{IB}}(\mathbf{x}_i^{\text{image}})
\end{split}
\end{equation}
To ensure that the outputs of the behavior-aligned encoder and the behavior-specific encoder are different, we minimize their mutual information by CLUB-based mutual information minimization. Compared to methods that try to optimize the lower limit of mutual information, such as InfoNCE\cite{infonce} and MINE\cite{mine}, CLUB\cite{club} can effectively optimize the upper limit of mutual information, demonstrating superior advantages in the disentanglement of information\cite{cross_modal}.
\begin{equation}
\begin{split}
\mathcal{L}_{MIM} = \sum_{m \in \{text,image\}} & \frac{1}{N_B} \sum_{i=1}^{N_B} \Bigg[  
    \log q_\theta\big(\overline{\mathbf{z}}_i^m \mid \mathbf{z}_i^m\big) \\
    - & \frac{1}{N_B} \sum_{j=1}^{N_B} \log q_\theta\big(\overline{\mathbf{z}}_j^m \mid \mathbf{z}_i^m\big) 
    \vphantom{\sum_{j=1}^{N_B}} \Bigg]
\end{split}
\end{equation}
where $m \in \{text, image\}$ and $q_{\theta}$ is the variational approximation of the ground-truth posterior of $\overline{\mathbf{z}}_i^m$ given $\mathbf{z}_i^m$, and can be parameterized by a network $\theta$. However, merely minimizing mutual information is insufficient to ensure that the results are meaningful and behavior-aligned. Therefore, we further propose a behavior-aligned quantization strategy to address this issue.

\subsubsection{Behavior Quantization Strategy}
To ensure that the behavior information is disentangled, our objective is to mapping the behavior-aligned information in a latent space dominated by behavior. Therefore, we designed a quantization strategy in which different modalities and behaviors share a unified codebook during quantization. This shared codebook serves as a bridge between information from different modalities and behaviors, facilitating implicit alignment.

Inspired by previous work\cite{id_vs_modal,missrec}, we recognize that only modality information cannot encompass all information in the behavior pattern. Therefore, we supplement the item with an additional ID representation, denoted as \(\mathbf{z}_i^{\text{ID}}\), which is learnable and initialized randomly as supplementary. Consequently, the complete behavior representation will obtain three components:
\begin{align}
    \mathbf{Z}_i = \{\mathbf{z}_i^{\text{ID}}, \mathbf{z}_i^{\text{text}}, \mathbf{z}_i^{\text{image}}\}
\end{align}
Each information is useful to capture the user intent by enriching the item's information through complementary aspects and we will discuss them in Section \ref{ref:bs}. Due to the widespread application of RQ-VAE in generative recommendation\cite{tiger,letter,cost}, we introduce it as our vector quantized method to represent these three types of behavioral information.

Residual quantization is an iterative process that quantizes residual embedding at each step, producing a sequence of semantic IDs. Starting with an input latent representation $\mathbf{z}_m \in \mathbf{Z}_i$ and a codebook $\mathcal{C}^l$ containing $N$ codes of dimension $D$, we identify the closest codebook vector at each quantization level. The process begins with $\mathbf{z}^0_m = \mathbf{z}_m$, and the closest vector from the codebook is determined using equation \ref{eq:rq_idx}.
\begin{equation}\label{eq:rq_idx}
\mathbf{c}_m^l = \arg\min_{\mathbf{c} \in \mathcal{C}^l} \| \mathbf{z}_m^l - \mathbf{c} \|^2
\end{equation}
Once the representation for this layer is obtained, the residual representation will then be computed and propagated to the next level.
\begin{equation}\label{eq:rq_res}
\mathbf{z}_m^{l+1} = \mathbf{z}_m^l - \mathbf{c}_m^l
\end{equation}
Iterating the process in Equations \ref{eq:rq_idx} and \ref{eq:rq_res} for $L$ times, the final quantized representation $\mathbf{q}^m_i$ is the sum of all selected codebook vectors.
\begin{equation}
\mathbf{q}^m_i = \sum_{l=0}^{L} \mathbf{c}_m^l
\end{equation}
Subsequently, the decoder $\mathbf{Decoder_{TR}}$ and $\mathbf{Decoder_{IR}}$ will reconstruct the input from the quantized representation. Since we have disentangled the original modality information into modality-specific and behavior-aligned representations, both types of information are fed into the decoder:
\begin{equation}
\begin{split}
\hat{x}^{\text{text}}_i &= \mathbf{Decoder}_{TR}(\mathbf{q}^{\text{text}}_i \oplus \overline{\mathbf{z}}_i^{\text{text}}) \\
\hat{x}^{\text{image}}_i &= \mathbf{Decoder}_{IR}(\mathbf{q}^{\text{image}}_i \oplus \overline{\mathbf{z}}_i^{\text{image}})
\end{split}
\end{equation}
After this, we can use the reconstruction loss $\mathcal{L}_{\text{recon}}$ to ensure that the encodings retain the essential information of the original input. On the other hand, the loss of RQ $\mathcal{L}^{m}_{\text{RQ}}$ in quantization helps to ensure proper gradient backpropagation by minimizing the difference between the encoded representation and its quantized results.
\begin{equation}
\begin{split}
\mathcal{L}_{\text{recon}} = & \big\|\mathbf{x}_i^{\text{text}} - \mathbf{\hat{x}}_i^{\text{text}}\big\|_2^2 + \big\|\mathbf{x}_i^{\text{image}} - \mathbf{\hat{x}}_i^{\text{image}}\big\|_2^2 \\
\mathcal{L}^{m}_{\text{RQ}} = & \sum_{l=1}^{L} \left( \left\|\operatorname{sg}\left[\mathbf{z}_m^l\right] - \mathbf{c}_m^l \right\|_2^2 + \alpha^m \left\|\mathbf{z}_m^l - \operatorname{sg}[\mathbf{c}_m^l]\right\|_2^2 \right) \\
\mathcal{L}_{\text{RQ-VAE}} =& \mathcal{L}_{\text{recon}} + \sum_{m \in \{\text{text}, \text{image}, \text{ID}\}} \mathcal{L}^{m}_{\text{RQ}}
\end{split}
\end{equation}
where $m \in \{ID, \text{text}, \text{image}\}$, $sg[\cdot]$ is the stop gradient operation\cite{vq}, and $\alpha^m$ is the coefficient to balance the strength between the optimization of the code embedding and the modality-aligned encoder. Subsequently, each item is discretized into $3 \times L$ codes. Inspired by PQ \cite{pq}, we concatenate these three distinct signals into a single semantic ID, as they are parallel in nature and this concatenation will reduce the length of the prediction.

\subsubsection{Sequence-to-Item Contrastive Task}
To ensure that the behavioral signals are properly disentangled, we subsequently designed a sequence-to-item contrastive task to utilize these signals to learn a behavior recommendation task. We first concatenate these behavior representations from $\mathbf{Z}_i$ and obtain a comprehensive feature vector $\mathbf{v}_i$ for each item:
\begin{equation}
\mathbf{v}_i = \mathbf{q}_i^{\text{ID}} \oplus \mathbf{q}_i^{\text{text}} \oplus \mathbf{q}_i^{\text{image}}
\end{equation}
The operation $\oplus$ denotes concatenation. After this, we can use a sequence encoder, such as SASRec or SRGNN, to compress the user's historical behavior to $\mathbf{h}_t$
\begin{equation}
\mathbf{h}_t = f(\mathbf{v}_1, \mathbf{v}_2, \cdots, \mathbf{v}_t)
\end{equation}
Since the sequence representation leverages these quantized item embedding, we then predict next item by a contrastive loss as follow:
\begin{equation}
\mathcal{L}_{\text{Rec}} = -\log \frac{e^{\cos(\mathbf{h}_t, \mathbf{v}^+_{t+1}) / \tau}}{e^{\cos(\mathbf{h}_t, \mathbf{v}^+_{t+1}) / \tau} + \sum_{i \in B - \{i^+_{t+1}\}} e^{\cos(\mathbf{h}_t, \mathbf{v}^-_i) / \tau}}
\end{equation}
where \(\cos(\mathbf{h}_t, \mathbf{v}_i)\) represents the cosine similarity between the sequence representation $\mathbf{h}_t$ and the item representation \(\mathbf{v}_i\). \(\tau\) is the temperature parameter that scales the logits. $\mathbf{B}$ is the batch size. This sequence-to-item contrastive task leverages the direct optimization of signals in the behavioral domain, thereby ensuring that the representations are behavior-aligned.

\subsubsection{overall loss}We use a weighted sum of loss terms to flexibly integrate multiple task objectives in the quantization process. The overall loss function is as follows:
\begin{equation}
\mathcal{L}_{\text{total}} = \mathcal{L}_{\text{RQ-VAE}} + \beta \mathcal{L}_{\text{Rec}} + \gamma \mathcal{L}_{\text{MIM}}
\end{equation}
where $\beta$ and $\gamma$ are hyperparameters used to control weights between different losses.
\subsection{Similarity-enhanced Self-attention}
We have made significant efforts on the quantization representation stage and also hope to take advantage of these representations to perform downstream tasks better. However, directly inheriting these representations has been shown in many studies to limit the performance of recommendation systems\cite{nova,difsr,asif}. Inspired by these works, we propose a non-invasive method to leverage the semantic representations obtained during the quantization process. Specifically, for each pair of items $(i, j)$ in the sequence, we first calculate the similarity between their quantized representation:
\begin{equation}
d_{ij} = \frac{\mathbf{v}_i \cdot \mathbf{v}_j}{\|\mathbf{v}_i\| \|\mathbf{q}_j\|}
\end{equation}
Here, $\mathbf{v}_i$ and $\mathbf{v}_j$ are the representations of the codebook after completing the quantization training task. The cosine similarity $d_{ij}$ lies in the range $[-1, 1]$. We then apply the following transformation to discretize it into an integer in the range $[0, K]$:
\begin{equation}
\text{sim}_{ij} = \text{Round}\left(\frac{d_{ij} + 1}{2} \times K\right)
\end{equation}
Subsequently, we use the index $\text{sim}_{ij}$ to retrieve the corresponding similarity representation $\mathbf{e}^{\text{sim}}_{ij}$ from a similarity embedding table of length $K$. This representation is then seamlessly integrated into the attention mechanism. The traditional self-attention mechanism is described as follows:
\begin{equation}
\begin{split}
\alpha^{kl}_{ij} = \frac{\exp\left(\frac{(\mathbf{W}_Q \cdot \mathbf{e}^k_i) \cdot (\mathbf{W}_K \cdot \mathbf{e}^l_j)}{\sqrt{D}}\right)}{\sum_{j=1}^t \sum_{l=1}^L \exp\left(\frac{(\mathbf{W}_Q \cdot \mathbf{e}^k_i) \cdot (\mathbf{W}_K \cdot \mathbf{e}^l_n)}{\sqrt{D}}\right)}
\end{split}
\end{equation}
where $W_Q, W_K, W_V$ are parameter matrices mapping elements to the query, key, and value spaces. $\mathbf{e}^k_i, \mathbf{e}^l_j$ is the embedding of the item's semantic IDs, which are initialized randomly. After this, we use the following method to complete this non-invasive integration strategy:
\begin{equation}
\begin{split}
\text{Sim-Attention}(\mathbf{e}_i^k, \mathbf{e}_j^l) = 
\mathbf{e}^{\text{sim}}_{ij} \odot \underbrace{ \left( \alpha^{kl}_{ij} \times \left(\mathbf{W}_V \cdot \mathbf{e}^l_{j}\right) \right) }_{\text{traditional self-attention}}
\end{split}
\end{equation}
where $\odot$ denotes the bitwise multiplication operation. This approach does not directly inherit the semantic representation. Instead, it quantifies the similarity of semantic representations and applies it in a bitwise manner to the value, which ensures that the quantized representation of semantic IDs does not harm the learning of IDs.

\begin{table}[h]
\caption{Statistics of datasets used for experiments (a.v.l = average sequence length).}\label{t:dataset}
\begin{tabular}{ccccc}
\toprule
Dataset       & Beauty  & Sports & Clothing & Toys    \\ 
\midrule
\# Users      & 22,363  & 35,598      & 39,387   & 19,412   \\
\# Items      & 12,101  & 18,357      & 23,033   & 11,924   \\
\# Interactions & 198,502 & 296,337    & 278,677  & 169,597 \\
\# Images      & 12,009  & 18,177      & 22,879   & 11,838    \\
\# Texts       & 12,094  & 18,317      & 23,032   & 11,890    \\ 
a.v.l         & 8.88    & 8.32        & 7.08     & 8.63     \\
\bottomrule
\end{tabular}
\end{table}
\section{Experiment}
In this section, we evaluate \baby by comparing it with traditional, modality-fused, and vector-quantized sequential recommenders. We begin by introducing the experimental setup and analyze the experimental results.
\subsection{Experimental Setup}
\begin{table*}[htbp]
\centering
\caption{The performance comparison between the baselines and \baby is presented. The best performance in each row is highlighted in bold, while the second-best is underlined. The symbol $\blacktriangle$ indicates the relative improvement of our results over the best baseline, which is consistently significant at the 0.05 level.}
\setlength{\tabcolsep}{1.5pt} 
\begin{tabular}{c|l|ccc|cccc|cccc|c}
\toprule
\multirow{2}{*}{Dataset} & \multicolumn{1}{c|}{\multirow{2}{*}{Metric}} & \multicolumn{3}{c|}{Traditional} & \multicolumn{4}{c|}{Modality-fused} & \multicolumn{4}{c|}{Vector-quantized} & \multirow{2}{*}{$\blacktriangle \%$} \\
                         & \multicolumn{1}{c|}{}                         & GRU4Rec & BERT4Rec & SASRec & $\rm{GRU4Rec}_F$ & $\rm{SASRec}_F$ & NOVA-BERT & MISSRec & VQ-Rec & TIGER & LETTER & \baby \\ 
\midrule
\multirow{4}{*}{Beauty} & R@5  & 0.0475 & 0.0602 & 0.0641 & 0.0537 & 0.0662 & 0.0708 & 0.0749 & 0.0728 & 0.0742 & \underline{0.0750} & \textbf{0.0815} & +8.67\% \\
                        & R@10 & 0.0727 & 0.0856 & 0.0885 & 0.0800 & 0.0933 & 0.1003 & 0.1024 & 0.1012 & 0.1039 & \underline{0.1063} & \textbf{0.1147} & +7.90\% \\
                        & N@5  & 0.0327 & 0.0417 & 0.0466 & 0.0373 & 0.0483 & 0.0513 & \underline{0.0538} & 0.0527 & 0.0534 & 0.0534 & \textbf{0.0576} & +7.06\% \\
                        & N@10 & 0.0408 & 0.0498 & 0.0544 & 0.0457 & 0.0570 & 0.0608 & 0.0626 & 0.0618 & 0.0629 & \underline{0.0642} & \textbf{0.0683} & +6.38\% \\ 
\hline
\multirow{4}{*}{Sports} & R@5  & 0.0316 & 0.0318 & 0.0353 & 0.0330 & 0.0403 & 0.0411 & 0.0421 & 0.0434 & 0.0450 & \underline{0.0467} & \textbf{0.0501} & +7.28\%\\
                        & R@10 & 0.0480 & 0.0506 & 0.0499 & 0.0508 & 0.0545 & 0.0570 & 0.0587 & 0.0620 & 0.0643 & \underline{0.0664} & \textbf{0.0721} & +8.58\% \\
                        & N@5  & 0.0215 & 0.0212 & 0.0250 & 0.0220 & 0.0286 & 0.0287 & 0.0296 & 0.0300 & 0.0312 & \underline{0.0323} & \textbf{0.0344} & +6.50\% \\
                        & N@10 & 0.0267 & 0.0273 & 0.0297 & 0.0278 & 0.0332 & 0.0353 & 0.0350 & 0.0360 & 0.0374 & \underline{0.0386} & \textbf{0.0415} & +7.51\% \\ 
\hline
\multirow{4}{*}{Clothing} & R@5  & 0.0166 & 0.0182 & 0.0204 & 0.0182 & 0.0262 & 0.0287 & \underline{0.0361} & 0.0298 & 0.0323 & 0.0345 & \textbf{0.0402} & +11.36\% \\
                          & R@10 & 0.0252 & 0.0291 & 0.0278 & 0.0297 & 0.0383 & 0.0415 & \underline{0.0533} & 0.0428 & 0.0482 & 0.0528 & \textbf{0.0589} & +10.51\% \\
                          & N@5  & 0.0109 & 0.0120 & 0.0138 & 0.0117 & 0.0181 & 0.0205 & \underline{0.0246} & 0.0200 & 0.0218 & 0.0250 & \textbf{0.0272} & +10.56\% \\
                          & N@10 & 0.0137 & 0.0155 & 0.0162 & 0.0154 & 0.0220 & 0.0254 & 0.0301 & 0.0242 & 0.0270 & \underline{0.0306} & \textbf{0.0336} & +9.80\% \\ 
\hline
\multirow{4}{*}{Toys} & R@5  & 0.0506 & 0.0632 & 0.0691 & 0.0572 & 0.0772 & 0.0826 & 0.0854 & 0.0829 & 0.0854 & \underline{0.0864} & \textbf{0.0968} & +12.04\% \\
                      & R@10 & 0.0727 & 0.0863 & 0.0911 & 0.0810 & 0.1039 & 0.1169 & 0.1164 & 0.1117 & 0.1179 & \underline{0.1189} & \textbf{0.1299} & +9.25\% \\
                      & N@5  & 0.0363 & 0.0447 & 0.0502 & 0.0402 & 0.0573 & 0.0575 & 0.0612 & 0.0597 & 0.0620 & \underline{0.0627} & \textbf{0.0693} & 10.53\% \\
                      & N@10 & 0.0434 & 0.0522 & 0.0573 & 0.0479 & 0.0659 & 0.0687 & 0.0713 & 0.0690 & 0.0725 & \underline{0.0731} & \textbf{0.0800} & 9.44\% \\ 
\bottomrule
\end{tabular}\label{t:main}
\end{table*}
\paratitle{Dataset.}  
We conduct our experiments on the commonly used Amazon dataset with respect to its rich auxiliary information and used by many works \cite{vqrec,unisrec,tiger}.
To analyze the capability of our model, we select four different categories, including \textit{Beauty, Clothing\_Shoes\_and\_Jewelry, Toys\_and\_Games and Sports\_and\_Outdoors}.
For these categories, we remove users and items with fewer than 5 related actions and limit the number of items in a user’s history to 20 which is a common usage in many research works\cite{letter,tiger,eager}. The statistics of four datasets are shown in Table~\ref{t:dataset}. 

\paratitle{Baselines.}To evaluate the effectiveness of our approach, We compare \baby against three types of baselines, including three traditional sequential models, four modality-fused sequential models and three vector-quantized sequential models. The traditional sequential-based models include:
\begin{itemize}
\item \textbf{GRU4Rec}\cite{gru}: GRU4Rec is a session-based recommendation, which utilizes GRU unit to capture users' long sequential behaviors for recommendation.
\item  \textbf{SASRec}\cite{sasrec}: SASRec is a self-attention based sequential recommendation model that employs a causal decoder to predict the next item.
\item \textbf{BERT4Rec}\cite{bert4rec}: BERT4REc is a model that leverages the BERT architecture to enhance sequential recommendation tasks.
\end{itemize}

For modality-fused sequential models, we consider the following four baselines:
\begin{itemize}
\item $\rm{\mathbf{GRU4Rec_F}}$~\cite{gruf}: $\rm{GRU4Rec}_F$ integrates modality information into GRU networks to enhance sequential recommendations. We combine the pre-trained modality representation with item vectors to serve as the input for the GRU.
\item $\rm{\mathbf{SASRec_F}}$: Similarly to $\rm{GRU4Rec}_F$, we extend SASRec with the concatenation of item embeddings and the pre-trained modality representations.
\item \textbf{NOVA-BERT}\cite{nova}: NOVA-BERT uses a non-invasive self-attention mechanism to utilize the side information.
\item \textbf{MISSRec}\cite{missrec}: MISSRec is a
novel multi-modal pre-training and transfer learning framework for sequential recommendation.
\end{itemize}
For vector-quantized sequential models, we compare our approach with the following baselines:
\begin{itemize}
\item \textbf{VQ-Rec}\cite{vqrec}: VQ-Rec uses product quantization\cite{pq} to generate codes using textual information for item representation. 
\item \textbf{TIGER}\cite{tiger}: TIGER is a generative model which uses RQ-VAE\cite{rqvae} to generate Semantic IDs, leading to the hierarchical representation of items.
\item \textbf{LETTER}\cite{letter}: LETTER integrates hierarchical semantics, collaborative signals, and code assignment diversity to enhance the quality of the Semantic IDs. In our paper, we use the version of LETTER-TIGER as our baseline.
\end{itemize}

\paratitle{Evaluation Metric}. To present a comprehensive evaluation, we sort user's records by timestamp to create the interaction sequence. Based on the sorted sequences, we hold out the last item of each sequence as the test data and the second-to-last item for validation. The remaining data is treated as the training data.

We use Recall@k~(R@k), NDCG@k(N@k) as our evaluation metrics(k=5/10). R$@k$ measures the percentage of target items appearing in the top-$k$ results, and
N$@k$ takes the ranking position in the top-$k$ list into account. 
We perform significant tests using the paired t-test. Differences are considered statistically significant when the $p$-value is lower than $0.05$.

\paratitle{Experiment Settings}. 
For fair comparison, we adopt the following settings for all methods: the batch size is set to 256; all embedding parameters are randomly initialized in the range of (0, 0.01); 
the model dimension is tuned in the range of$[32, 64, 96, 128, 256]$. For VQ-Rec\footnote{https://github.com/RUCAIBox/VQ-Rec} and MISSRec\footnote{https://github.com/gimpong/MM23-MISSRec} and LETTER\footnote{https://github.com/HonghuiBao2000/LETTER}, we use the code relased by the author. For other baselines, we implement them based on RecBole~\cite{recbole}. We optimize all the methods according to the validation sets. For textual information, we concatenate both the title and description of the items and obtain the embedding from a pre-trained sentence T5~\cite{sentence-t5};
for visual information, we use the embedding provided by amazon \cite{amazon, amazon_1}. For our model, both the weight $\beta$ of $\mathcal{L}_{Rec}$  and the weight $\gamma$ of $\mathcal{L}_{MIM}$ is set to 1. Folowed by \cite{unisrec,letter}, the temperature coefficient $\tau$ is set to 0.1 and $\alpha^m$ in $\mathcal{L}^m_{\text{RQ}}$ is set to 0.25 for each modality.
\begin{figure*}[h]
\centering
\includegraphics[scale=0.31]{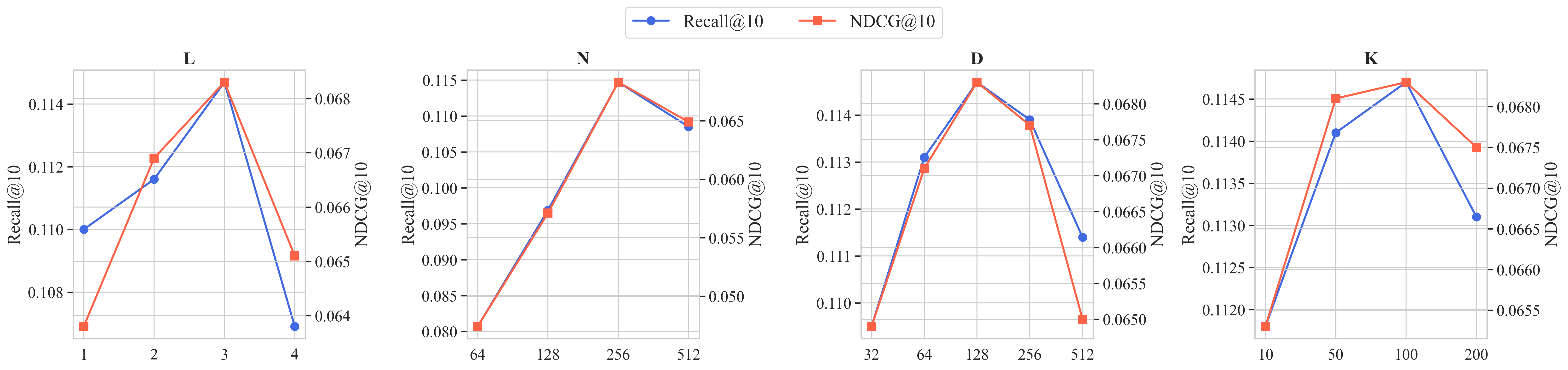}
\caption{Performance of \baby on the Beauty Dataset with Different hyper-parameters.}
\label{fig:hyper}
\end{figure*}
\subsection{Performance Comparison}
In this section, we compare the performance of our model with the
baselines. The overall performance of our proposed \baby and the baselines are reported in Table~\ref{t:main}. We have the following observations:

For sequential recommendations, GRU4Rec obtains the worst performance. 
Compared to GRU4Rec, we found that both self-attention enhanced models ~(SASRec and BERT4Rec) can improve the performance of the recommendation.

After introducing the modality information, both $\rm{GRU4Rec}_F$ and $\rm{SASRec}_F$ perform better than their initial models GRU4Rec and SASRec. This demonstrates the effectiveness of fusing modality information to improve recommendation performance.  
Compared with modality-fused $\rm{GRU4Rec}_F$ and $\rm{SASRec}_F$, NOVA-BERT uses a non-invasive mechanism to better utilize multi-modal information, and MISSRec employs more complex modality-fusion strategies to align user behavior. Both models achieve better performance.
\begin{table}[h]
\small
\centering
\caption{Performance comparison of \baby and its three variants over four datasets. The best performance is in bold font.}
\setlength{\tabcolsep}{1.5pt} 
\label{t:quant}
\begin{tabular}{c|l|cccc}
\hline
Dataset & Metrics & {$\rm{\baby}_{\neg{MIM}}$} & {$\rm{\baby}_{\neg{Rec}}$} & \multicolumn{1}{c}{$\rm{\baby}_{\neg{U}}$} & \baby \\ \hline
\multirow{4}{*}{Beauty} & R@5  & 0.0788 & 0.0755 & 0.0764  & \textbf{0.0815} \\
                        & R@10 & 0.1127 & 0.1061 & 0.1109  & \textbf{0.1147} \\
                        & N@5    & 0.0556 & 0.0529 & 0.0537  & \textbf{0.0576} \\
                        & N@10   & 0.0660 & 0.0638 & 0.0648  & \textbf{0.0683} \\ \hline
\multirow{4}{*}{Sports} & R@5  & 0.0487 & 0.0469 & 0.0473  & \textbf{0.0501} \\
                            & R@10 & 0.0683 & 0.0663 & 0.0683  & \textbf{0.0721} \\
                            & N@5    & 0.0331 & 0.0321 & 0.0333  & \textbf{0.0344} \\
                            & N@10   & 0.0395 & 0.0383 & 0.0391  & \textbf{0.0415} \\ \hline
\multirow{4}{*}{Clothing} & R@5  & 0.0386 & 0.0340 & 0.0379  & \textbf{0.0402} \\
                          & R@10 & 0.0580 & 0.0533 & 0.0577  & \textbf{0.0589} \\
                          & N@5    & 0.0262 & 0.0246 & 0.0259  & \textbf{0.0272} \\
                          & N@10   & 0.0321 & 0.0305 & 0.0320  & \textbf{0.0336} \\ \hline
\multirow{4}{*}{Toys} & R@5  & 0.0947 & 0.0875 & 0.0935  & \textbf{0.0968} \\
                        & R@10 & 0.1280 & 0.1216 & 0.1274  & \textbf{0.1299} \\
                        & N@5    & 0.0670 & 0.0626 & 0.0656  & \textbf{0.0693} \\
                        & N@10   & 0.0790 & 0.0736 & 0.0778  & \textbf{0.0800} \\ \hline
\end{tabular}\label{tb:ab_quant}
\end{table}

\begin{figure*}[htbp]
\centering
\includegraphics[scale=0.23]{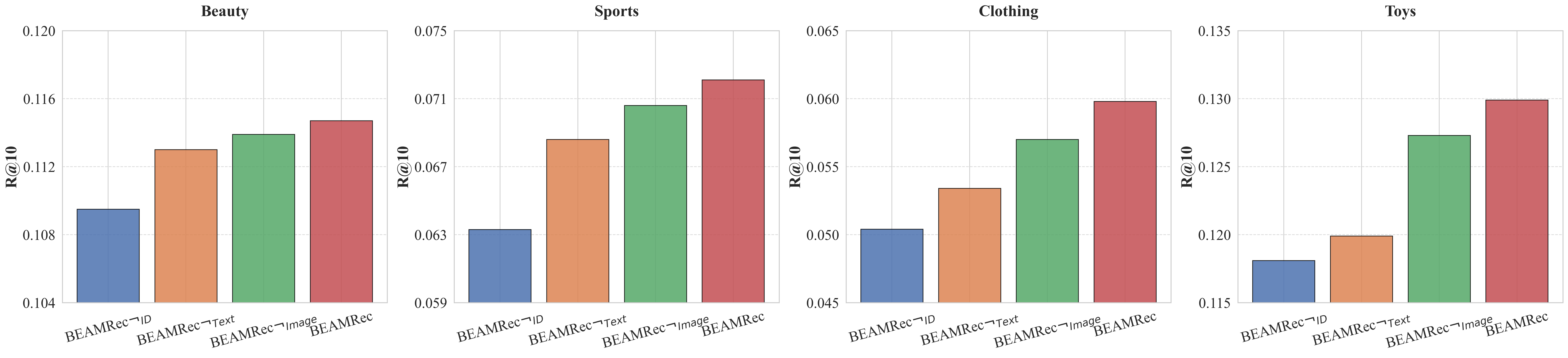}
\caption{The results of the three variants of \baby, namely $\baby_{\neg{\text{ID}}}$, $\baby_{\neg{\text{text}}}$, and $\baby_{\neg{\text{image}}}$, on four datasets.}
\label{fig:modal_analysis}
\end{figure*}
For vector-quantized sequential recommendation, we found that VQ-Rec achieved the poorest performance due to its lack of generative capability and its primary use in transfer recommendation. TIGER uses RQ-VAE to generate hierarchical semantic IDs and autoregressively generate items, achieving better performance. LETTER further refined the quantization module, incorporating signals such as behavioral signals, resulting in improved results. Finally, our proposed approach, \baby, achieves the best performance among all methods on four datasets. Our experimental results also indicate that a well-designed quantization strategy and the effective use of quantized representations can further enhance the performance of the downstream generative recommendation task.
\subsection{Ablation Study}
In this section, we design experiments to consider the impact of different variants of \baby on the results.

\subsubsection{Analysis on Behavior-aligned Quantized Strategy}
In this section, we want to analyze whether such a design can bring benefits. We discuss the three variants of \baby as follows.
\begin{itemize}
    \item $\rm{\baby}_{\neg{MIM}}$: we remove the $\mathcal{L}_{MIM}$ and the corresponding modality-specific encoder is also removed, as these two components work together.
    \item $\rm{\baby}_{\neg{Rec}}$: we remove the $\mathcal{L}_{Rec}$ and the additional ID representation, as this representation is only supervised in the sequence-to-item contrastive task.
    \item $\rm{\baby}_{\neg{U}}$: we remove the shared codebook, thereby each behavior signal being quantized independently.
\end{itemize}
As indicated in Table \ref{tb:ab_quant}, eliminating the MIM loss results in a decline in performance which demonstrate that directly mapping modality information into the codebook will introduces noise. Similarly, removing the recommendation loss leads to a marked decrease in performance, confirming the necessity of incorporating supervised behavior signals to ensure the disentanglement of behavior-aligned information. Furthermore, eliminating the shared codebook results in a significant drop in performance, supporting our assertion that a shared codebook allows modalities to implicitly align with behaviors, thereby enhancing the success of collaborative tasks.

\subsubsection{Analysis on Similarity-enhanced Self-attention}
We validate \baby's non-invasive strategy through two ablation studies. 
\begin{itemize}
\item $\baby_{S}$: we remove the auxiliary module in self-attention module, which means the quantized representation will be dropped. 
\item $\baby_{E}$: we inherit the semantic ID representation obtained by the item tokenization stage and makes them learnable in the generation stage.
\end{itemize}
As shown in Table~\ref{t:sim}, removing the similarity module ($\baby_{S}$) causes significant performance drops across all datasets, validating its critical role. Furthermore, semantic ID inheritance ($\baby_{E}$) not only fails to enhance performance but introduces marginal degradation.

\begin{table}[htbp]
\centering
\caption{Performance comparison of \baby, $\baby_{S}$, and $\baby_{E}$ over four datasets. The best performance is in bold font.}
\label{t:sim}
\begin{tabular}{c|l|ccc}
\hline
Dataset & Metrics & {$\rm{\baby}_{S}$} & \multicolumn{1}{c}{$\rm{\baby}_{E}$} & \baby \\ \hline
\multirow{4}{*}{Beauty} & R@5  & 0.0778 & 0.0768  & \textbf{0.0815} \\
                        & R@10 & 0.1093 & 0.1075  & \textbf{0.1147} \\
                        & N@5    & 0.0541 & 0.0535  & \textbf{0.0576} \\
                        & N@10   & 0.0642 & 0.0634  & \textbf{0.0683} \\ \hline
\multirow{4}{*}{Sports} & R@5  & 0.0478 & 0.0442  & \textbf{0.0501} \\
                        & R@10 & 0.0678 & 0.0637  & \textbf{0.0721} \\
                        & N@5    & 0.0333 & 0.0307  & \textbf{0.0344} \\
                        & N@10   & 0.0397 & 0.0370  & \textbf{0.0415} \\ \hline
\multirow{4}{*}{Clothing} & R@5  & 0.0384 & 0.0345  & \textbf{0.0402} \\
                          & R@10 & 0.0577 & 0.0525  & \textbf{0.0598} \\
                          & N@5    & 0.0261 & 0.0232  & \textbf{0.0272} \\
                          & N@10   & 0.0323 & 0.0289  & \textbf{0.0336} \\ \hline
\multirow{4}{*}{Toys} & R@5  & 0.0917 & 0.0869  & \textbf{0.0968} \\
                      & R@10 & 0.1269 & 0.1228  & \textbf{0.1299} \\
                      & N@5    & 0.0639 & 0.0595  & \textbf{0.0693} \\
                      & N@10   & 0.0752 & 0.0710  & \textbf{0.0800} \\ \hline
\end{tabular}
\end{table}
\subsection{Effectiveness of the Hyper-parameter}
We investigated how various hyper-parameter settings affect the model's performance on the Beauty dataset. The baseline parameters are set as follows: $L=3$, $N=256$, $D=128$, and $K=100$. In each analysis, specific parameters were adjusted, while the others remained constant.

\paratitle{Impact of Semantic ID Length $\textbf{L}$.}
As shown in Figure \ref{fig:hyper}. We observed a significant improvement when increasing the length from 1 to 3, indicating that hierarchical semantic IDs in RQ offer better performance when compared to VQ\cite{vq}. A noticeable decline in performance was observed when the length increased to 4 due to the increased difficulty in predicting more codes and the impact of cumulative errors in autoregressive models.

\paratitle{Impact of Semantic Codebook Size $\textbf{N}$.}
Figure~\ref{fig:hyper} demonstrates that performance improved as the size of the codebook increased from 64 to 256. However, when the size of the codebook reached 512, the performance began to decline. This may be due to the increased number of cluster centers causing the target to become sparse, reducing the model's generalization ability and increasing the risk of overfitting.

\paratitle{Impact of Semantic Codebook Dimension $\textbf{D}$.}
As shown in Figure \ref{fig:hyper}. We found that before the dimension reaches 128, the model's performance improves as the embedding dimension of the codebook increases; however, once the dimension exceeds 128, the performance drops rapidly. This indicates that selecting an appropriate codebook dimension also affects the quality of quantization.

\paratitle{Impact of Similarity Bucket Size $\textbf{K}$.}
Figure~\ref{fig:hyper} reveals an initial performance improvement as similarity buckets increase from 10 to 100, but slightly decreased at 200 buckets. This non-monotonic pattern suggests balanced granularity control sufficient buckets enable fine-grained distinction (100-bucket optimal), while excessive quantization amplifies similarity measurement noise, triggering model overfitting.
\subsection{Effectiveness of Component of Behavior Singnls}\label{ref:bs}
In \baby, we evaluate the impact of three behavioral signal components (ID embeddings, behavior-aligned text embeddings, and behavior-aligned image embeddings) by creating ablated variants: removes learnable ID embeddings ($\baby_{\neg{\text{ID}}}$), excludes pre-trained text features ($\baby_{\neg{\text{text}}}$), and eliminates visual embeddings ($\baby_{\neg{\text{image}}}$). each variant disables corresponding modules associated with the removed component.

Figure \ref{fig:modal_analysis} demonstrates the indispensable nature of all behavioral information types through cross-dataset validity: 1) Behavioral signal removal causes the most significant performance degradation, confirming modality information's auxiliary but non-reconstructive role in behavioral learning; 2) Text modality removal impacts performance more substantially than image removal; 3) \baby's best performance with full integration of all three information types validates effective multi-modal collaboration.
\subsection{Further Analysis}
\subsubsection{Item Collisions}
Since we use a series of semantic IDs to represent items, whether an item has unique semantic IDs sequence affects the capability of generative models\cite{tiger}. 
\begin{table}[htbp]
\centering
\caption{The results show the number of items with the same semantic IDs sequence in the Toys and Clothing datasets.}
\label{t:quant}
\begin{tabular}{c|c|ccc}
\hline
Dataset & Length & RQ-VAE & LETTER & \baby \\ \hline
\multirow{4}{*}{Toys} & L=2 & 2154 & 2522  & 12 \\
                        & L=3   & 466 & 167  & 0 \\
                        & L=4   & 163 & 19  & 0 \\ \hline
\multirow{4}{*}{Clothing} & L=2 & 6014 & 6270  & 6 \\
                            & L=3 & 1173 & 395  & 1 \\
                            & L=4    & 771 & 23  & 0 \\ \hline
\end{tabular}
\end{table}
As shown in table~\ref{t:quant}, TIGER uses pure RQ-VAE quantization and achieves the highest conflict rate, while LETTER with diversified loss functions outperforms TIGER at codebook sizes $\geq 3$. \baby encodes three behavioral signals into one semantic ID, reducing conflicts without increasing the semantic IDs' length. This demonstrates our quantization strategy effectively mitigates codebook collisions without spatial complexity increases.
\subsubsection{Inference Speed Analysis}
In this section, we aim to test how our model's inference performance compares to the state-of-the-art generative model LETTER. Our inference performance is tested on a machine configured with Intel(R) Xeon(R) Platinum 8352Y CPU @ 2.20GHz $\times 2$  and NVIDIA A800-SXM4-80GB $\times$ 8 GPUs. We analyze the inference speed where topk=10 on Beauty dataset.
\begin{table}[h]
\centering
\caption{Analysis of the inference speed.}
\begin{tabular}{ccc}
\toprule
  & TIGER & \baby \\
\midrule
beamsize=50   & 0.010s   & 0.019s   \\
beamsize=100   & 0.015s   & 0.024s   \\
\bottomrule
\end{tabular} \label{tb:speed}
\end{table}

As shown in Table \ref{tb:speed}, despite our method involving the inference of three codes to form a semantic ID, there is no significant increase in inference time for a single sample. We attribute this to the fact that, although the search space is theoretically $O(N^3)$, the actual number of items is considerably smaller than it. Additionally, the high parallel processing capabilities of the GPU ensure that the model's inference time does not increase significantly.
\section{Conclusion}
We propose \baby, a generative sequential recommender addressing multi-modal information underutilization through two key innovations: 1) A behavior-aligned quantization strategy to decouple and quantize behavior-aligned multi-modal features from modality-specific noise. 2) A similarity-enhanced self-attention mechanism non-invasively incorporating quantized semantic relationships. Extensive experiments on four real-world datasets validate \baby's superior performance. In the future, it would be interesting to evaluate and deploy \baby on an online multimedia platform.

\newpage

\bibliographystyle{ACM-Reference-Format}
\bibliography{sample-base}


\begin{thebibliography}{60}


\ifx \showCODEN    \undefined \def \showCODEN     #1{\unskip}     \fi
\ifx \showDOI      \undefined \def \showDOI       #1{#1}\fi
\ifx \showISBNx    \undefined \def \showISBNx     #1{\unskip}     \fi
\ifx \showISBNxiii \undefined \def \showISBNxiii  #1{\unskip}     \fi
\ifx \showISSN     \undefined \def \showISSN      #1{\unskip}     \fi
\ifx \showLCCN     \undefined \def \showLCCN      #1{\unskip}     \fi
\ifx \shownote     \undefined \def \shownote      #1{#1}          \fi
\ifx \showarticletitle \undefined \def \showarticletitle #1{#1}   \fi
\ifx \showURL      \undefined \def \showURL       {\relax}        \fi
\providecommand\bibfield[2]{#2}
\providecommand\bibinfo[2]{#2}
\providecommand\natexlab[1]{#1}
\providecommand\showeprint[2][]{arXiv:#2}

\bibitem[Belghazi et~al\mbox{.}(2018)]%
        {mine}
\bibfield{author}{\bibinfo{person}{Mohamed~Ishmael Belghazi}, \bibinfo{person}{Aristide Baratin}, \bibinfo{person}{Sai Rajeswar}, \bibinfo{person}{Sherjil Ozair}, \bibinfo{person}{Yoshua Bengio}, \bibinfo{person}{R.~Devon Hjelm}, {and} \bibinfo{person}{Aaron~C. Courville}.} \bibinfo{year}{2018}\natexlab{}.
\newblock \showarticletitle{Mutual Information Neural Estimation}. In \bibinfo{booktitle}{\emph{ICML}} \emph{(\bibinfo{series}{Proceedings of Machine Learning Research}, Vol.~\bibinfo{volume}{80})}. \bibinfo{publisher}{{PMLR}}, \bibinfo{pages}{530--539}.
\newblock


\bibitem[Chang et~al\mbox{.}(2021)]%
        {dev_2}
\bibfield{author}{\bibinfo{person}{Jianxin Chang}, \bibinfo{person}{Chen Gao}, \bibinfo{person}{Yu Zheng}, \bibinfo{person}{Yiqun Hui}, \bibinfo{person}{Yanan Niu}, \bibinfo{person}{Yang Song}, \bibinfo{person}{Depeng Jin}, {and} \bibinfo{person}{Yong Li}.} \bibinfo{year}{2021}\natexlab{}.
\newblock \showarticletitle{Sequential Recommendation with Graph Neural Networks}. In \bibinfo{booktitle}{\emph{SIGIR}}. \bibinfo{pages}{378--387}.
\newblock


\bibitem[Cheng et~al\mbox{.}(2020)]%
        {club}
\bibfield{author}{\bibinfo{person}{Pengyu Cheng}, \bibinfo{person}{Weituo Hao}, \bibinfo{person}{Shuyang Dai}, \bibinfo{person}{Jiachang Liu}, \bibinfo{person}{Zhe Gan}, {and} \bibinfo{person}{Lawrence Carin}.} \bibinfo{year}{2020}\natexlab{}.
\newblock \showarticletitle{{CLUB:} {A} Contrastive Log-ratio Upper Bound of Mutual Information}. In \bibinfo{booktitle}{\emph{ICML}} \emph{(\bibinfo{series}{Proceedings of Machine Learning Research}, Vol.~\bibinfo{volume}{119})}. \bibinfo{publisher}{{PMLR}}, \bibinfo{pages}{1779--1788}.
\newblock


\bibitem[Choi et~al\mbox{.}(2021)]%
        {iim}
\bibfield{author}{\bibinfo{person}{Minjin Choi}, \bibinfo{person}{Jinhong Kim}, \bibinfo{person}{Joonseok Lee}, \bibinfo{person}{Hyunjung Shim}, {and} \bibinfo{person}{Jongwuk Lee}.} \bibinfo{year}{2021}\natexlab{}.
\newblock \showarticletitle{Session-aware Linear Item-Item Models for Session-based Recommendation}. In \bibinfo{booktitle}{\emph{{WWW}}}. \bibinfo{pages}{2186--2197}.
\newblock


\bibitem[de~Souza Pereira~Moreira et~al\mbox{.}(2021)]%
        {transformer4rec}
\bibfield{author}{\bibinfo{person}{Gabriel de Souza Pereira~Moreira}, \bibinfo{person}{Sara Rabhi}, \bibinfo{person}{Jeongmin Lee}, \bibinfo{person}{Ronay Ak}, {and} \bibinfo{person}{Even Oldridge}.} \bibinfo{year}{2021}\natexlab{}.
\newblock \showarticletitle{Transformers4Rec: Bridging the Gap between {NLP} and Sequential / Session-Based Recommendation}. In \bibinfo{booktitle}{\emph{RecSys}}. \bibinfo{publisher}{{ACM}}, \bibinfo{pages}{143--153}.
\newblock


\bibitem[Gao et~al\mbox{.}(2024)]%
        {gnr}
\bibfield{author}{\bibinfo{person}{Shen Gao}, \bibinfo{person}{Jiabao Fang}, \bibinfo{person}{Quan Tu}, \bibinfo{person}{Zhitao Yao}, \bibinfo{person}{Zhumin Chen}, \bibinfo{person}{Pengjie Ren}, {and} \bibinfo{person}{Zhaochun Ren}.} \bibinfo{year}{2024}\natexlab{}.
\newblock \showarticletitle{Generative News Recommendation}. In \bibinfo{booktitle}{\emph{WWW}}, \bibfield{editor}{\bibinfo{person}{Tat{-}Seng Chua}, \bibinfo{person}{Chong{-}Wah Ngo}, \bibinfo{person}{Ravi Kumar}, \bibinfo{person}{Hady~W. Lauw}, {and} \bibinfo{person}{Roy~Ka{-}Wei Lee}} (Eds.). \bibinfo{publisher}{{ACM}}, \bibinfo{pages}{3444--3453}.
\newblock


\bibitem[Ge et~al\mbox{.}(2016)]%
        {taper}
\bibfield{author}{\bibinfo{person}{Hancheng Ge}, \bibinfo{person}{James Caverlee}, {and} \bibinfo{person}{Haokai Lu}.} \bibinfo{year}{2016}\natexlab{}.
\newblock \showarticletitle{{TAPER:} {A} Contextual Tensor-Based Approach for Personalized Expert Recommendation}. In \bibinfo{booktitle}{\emph{RecSys}}. \bibinfo{pages}{261--268}.
\newblock


\bibitem[Geng et~al\mbox{.}(2022)]%
        {p5}
\bibfield{author}{\bibinfo{person}{Shijie Geng}, \bibinfo{person}{Shuchang Liu}, \bibinfo{person}{Zuohui Fu}, \bibinfo{person}{Yingqiang Ge}, {and} \bibinfo{person}{Yongfeng Zhang}.} \bibinfo{year}{2022}\natexlab{}.
\newblock \showarticletitle{Recommendation as Language Processing {(RLP):} {A} Unified Pretrain, Personalized Prompt {\&} Predict Paradigm {(P5)}}. In \bibinfo{booktitle}{\emph{RecSys}}. \bibinfo{publisher}{{ACM}}, \bibinfo{pages}{299--315}.
\newblock


\bibitem[Geng et~al\mbox{.}(2023)]%
        {vip5}
\bibfield{author}{\bibinfo{person}{Shijie Geng}, \bibinfo{person}{Juntao Tan}, \bibinfo{person}{Shuchang Liu}, \bibinfo{person}{Zuohui Fu}, {and} \bibinfo{person}{Yongfeng Zhang}.} \bibinfo{year}{2023}\natexlab{}.
\newblock \showarticletitle{{VIP5:} Towards Multimodal Foundation Models for Recommendation}. In \bibinfo{booktitle}{\emph{Findings of EMNLP}}. \bibinfo{publisher}{Association for Computational Linguistics}, \bibinfo{pages}{9606--9620}.
\newblock


\bibitem[He and McAuley(2016a)]%
        {amazon}
\bibfield{author}{\bibinfo{person}{Ruining He} {and} \bibinfo{person}{Julian McAuley}.} \bibinfo{year}{2016}\natexlab{a}.
\newblock \showarticletitle{Ups and downs: Modeling the visual evolution of fashion trends with one-class collaborative filtering}. In \bibinfo{booktitle}{\emph{WWW}}. \bibinfo{pages}{507--517}.
\newblock


\bibitem[He and McAuley(2016b)]%
        {vbpr}
\bibfield{author}{\bibinfo{person}{Ruining He} {and} \bibinfo{person}{Julian~J. McAuley}.} \bibinfo{year}{2016}\natexlab{b}.
\newblock \showarticletitle{{VBPR:} Visual Bayesian Personalized Ranking from Implicit Feedback}. In \bibinfo{booktitle}{\emph{AAAI}}. \bibinfo{pages}{144--150}.
\newblock


\bibitem[Hidasi et~al\mbox{.}(2016a)]%
        {gru}
\bibfield{author}{\bibinfo{person}{Balazs Hidasi}, \bibinfo{person}{Alexandros Karatzoglou}, \bibinfo{person}{Linas Baltrunas}, {and} \bibinfo{person}{Domonkos Tikk}.} \bibinfo{year}{2016}\natexlab{a}.
\newblock \showarticletitle{Session-based Recommendations with Recurrent Neural Networks}. In \bibinfo{booktitle}{\emph{ICLR}}.
\newblock


\bibitem[Hidasi et~al\mbox{.}(2016b)]%
        {gruf}
\bibfield{author}{\bibinfo{person}{Bal\'{a}zs Hidasi}, \bibinfo{person}{Massimo Quadrana}, \bibinfo{person}{Alexandros Karatzoglou}, {and} \bibinfo{person}{Domonkos Tikk}.} \bibinfo{year}{2016}\natexlab{b}.
\newblock \showarticletitle{Parallel Recurrent Neural Network Architectures for Feature-Rich Session-Based Recommendations}. In \bibinfo{booktitle}{\emph{RecSys}}. \bibinfo{pages}{241–248}.
\newblock


\bibitem[Hou et~al\mbox{.}(2023)]%
        {vqrec}
\bibfield{author}{\bibinfo{person}{Yupeng Hou}, \bibinfo{person}{Zhankui He}, \bibinfo{person}{Julian~J. McAuley}, {and} \bibinfo{person}{Wayne~Xin Zhao}.} \bibinfo{year}{2023}\natexlab{}.
\newblock \showarticletitle{Learning Vector-Quantized Item Representation for Transferable Sequential Recommenders}. In \bibinfo{booktitle}{\emph{WWW}}. \bibinfo{publisher}{{ACM}}, \bibinfo{pages}{1162--1171}.
\newblock


\bibitem[Hou et~al\mbox{.}(2022)]%
        {unisrec}
\bibfield{author}{\bibinfo{person}{Yupeng Hou}, \bibinfo{person}{Shanlei Mu}, \bibinfo{person}{Wayne~Xin Zhao}, \bibinfo{person}{Yaliang Li}, \bibinfo{person}{Bolin Ding}, {and} \bibinfo{person}{Ji{-}Rong Wen}.} \bibinfo{year}{2022}\natexlab{}.
\newblock \showarticletitle{Towards Universal Sequence Representation Learning for Recommender Systems}. In \bibinfo{booktitle}{\emph{KDD}}. \bibinfo{publisher}{{ACM}}, \bibinfo{pages}{585--593}.
\newblock


\bibitem[Huang et~al\mbox{.}(2018)]%
        {HuangZDWC18}
\bibfield{author}{\bibinfo{person}{Jin Huang}, \bibinfo{person}{Wayne~Xin Zhao}, \bibinfo{person}{Hong{-}Jian Dou}, \bibinfo{person}{Ji{-}Rong Wen}, {and} \bibinfo{person}{Edward~Y. Chang}.} \bibinfo{year}{2018}\natexlab{}.
\newblock \showarticletitle{Improving Sequential Recommendation with Knowledge-Enhanced Memory Networks}. In \bibinfo{booktitle}{\emph{SIGIR}}. \bibinfo{pages}{505--514}.
\newblock


\bibitem[J{\'{e}}gou et~al\mbox{.}(2011)]%
        {pq}
\bibfield{author}{\bibinfo{person}{Herv{\'{e}} J{\'{e}}gou}, \bibinfo{person}{Matthijs Douze}, {and} \bibinfo{person}{Cordelia Schmid}.} \bibinfo{year}{2011}\natexlab{}.
\newblock \showarticletitle{Product Quantization for Nearest Neighbor Search}.
\newblock \bibinfo{journal}{\emph{{IEEE} Trans. Pattern Anal. Mach. Intell.}} \bibinfo{volume}{33}, \bibinfo{number}{1} (\bibinfo{year}{2011}), \bibinfo{pages}{117--128}.
\newblock


\bibitem[Kang and McAuley(2018)]%
        {sasrec}
\bibfield{author}{\bibinfo{person}{Wang{-}Cheng Kang} {and} \bibinfo{person}{Julian~J. McAuley}.} \bibinfo{year}{2018}\natexlab{}.
\newblock \showarticletitle{Self-Attentive Sequential Recommendation}. In \bibinfo{booktitle}{\emph{ICDM}}. \bibinfo{pages}{197--206}.
\newblock


\bibitem[Kipf and Welling(2017)]%
        {gcn}
\bibfield{author}{\bibinfo{person}{Thomas~N. Kipf} {and} \bibinfo{person}{Max Welling}.} \bibinfo{year}{2017}\natexlab{}.
\newblock \showarticletitle{Semi-Supervised Classification with Graph Convolutional Networks}. In \bibinfo{booktitle}{\emph{ICLR}}. \bibinfo{publisher}{OpenReview.net}.
\newblock


\bibitem[Li et~al\mbox{.}(2019)]%
        {rns}
\bibfield{author}{\bibinfo{person}{Chenliang Li}, \bibinfo{person}{Xichuan Niu}, \bibinfo{person}{Xiangyang Luo}, \bibinfo{person}{Zhenzhong Chen}, {and} \bibinfo{person}{Cong Quan}.} \bibinfo{year}{2019}\natexlab{}.
\newblock \showarticletitle{A Review-Driven Neural Model for Sequential Recommendation}. In \bibinfo{booktitle}{\emph{IJCAI}}. \bibinfo{publisher}{International Joint Conferences on Artificial Intelligence Organization}, \bibinfo{pages}{2866--2872}.
\newblock


\bibitem[Li et~al\mbox{.}(2017)]%
        {narm}
\bibfield{author}{\bibinfo{person}{Jing Li}, \bibinfo{person}{Pengjie Ren}, \bibinfo{person}{Zhumin Chen}, \bibinfo{person}{Zhaochun Ren}, \bibinfo{person}{Tao Lian}, {and} \bibinfo{person}{Jun Ma}.} \bibinfo{year}{2017}\natexlab{}.
\newblock \showarticletitle{Neural Attentive Session-Based Recommendation}. In \bibinfo{booktitle}{\emph{CIKM}}. \bibinfo{pages}{1419–1428}.
\newblock


\bibitem[Li et~al\mbox{.}(2022b)]%
        {miner}
\bibfield{author}{\bibinfo{person}{Jian Li}, \bibinfo{person}{Jieming Zhu}, \bibinfo{person}{Qiwei Bi}, \bibinfo{person}{Guohao Cai}, \bibinfo{person}{Lifeng Shang}, \bibinfo{person}{Zhenhua Dong}, \bibinfo{person}{Xin Jiang}, {and} \bibinfo{person}{Qun Liu}.} \bibinfo{year}{2022}\natexlab{b}.
\newblock \showarticletitle{{MINER:} Multi-Interest Matching Network for News Recommendation}. In \bibinfo{booktitle}{\emph{Findings of ACL}}. \bibinfo{publisher}{Association for Computational Linguistics}, \bibinfo{pages}{343--352}.
\newblock


\bibitem[Li et~al\mbox{.}(2022a)]%
        {maris}
\bibfield{author}{\bibinfo{person}{Kaiyuan Li}, \bibinfo{person}{Pengfei Wang}, {and} \bibinfo{person}{Chenliang Li}.} \bibinfo{year}{2022}\natexlab{a}.
\newblock \showarticletitle{Multi-Agent RL-based Information Selection Model for Sequential Recommendation}. In \bibinfo{booktitle}{\emph{SIGIR}}. \bibinfo{publisher}{{ACM}}, \bibinfo{pages}{1622--1631}.
\newblock


\bibitem[Li et~al\mbox{.}(2020)]%
        {fashion_rec_1}
\bibfield{author}{\bibinfo{person}{Xingchen Li}, \bibinfo{person}{Xiang Wang}, \bibinfo{person}{Xiangnan He}, \bibinfo{person}{Long Chen}, \bibinfo{person}{Jun Xiao}, {and} \bibinfo{person}{Tat{-}Seng Chua}.} \bibinfo{year}{2020}\natexlab{}.
\newblock \showarticletitle{Hierarchical Fashion Graph Network for Personalized Outfit Recommendation}. In \bibinfo{booktitle}{\emph{SIGIR}}. \bibinfo{publisher}{{ACM}}, \bibinfo{pages}{159--168}.
\newblock


\bibitem[Liu et~al\mbox{.}(2021)]%
        {nova}
\bibfield{author}{\bibinfo{person}{Chang Liu}, \bibinfo{person}{Xiaoguang Li}, \bibinfo{person}{Guohao Cai}, \bibinfo{person}{Zhenhua Dong}, \bibinfo{person}{Hong Zhu}, {and} \bibinfo{person}{Lifeng Shang}.} \bibinfo{year}{2021}\natexlab{}.
\newblock \showarticletitle{Non-invasive Self-attention for Side Information Fusion in Sequential Recommendation}. In \bibinfo{booktitle}{\emph{AAAI}}. \bibinfo{pages}{4249--4256}.
\newblock


\bibitem[Liu et~al\mbox{.}(2024b)]%
        {mmgrec}
\bibfield{author}{\bibinfo{person}{Han Liu}, \bibinfo{person}{Yinwei Wei}, \bibinfo{person}{Xuemeng Song}, \bibinfo{person}{Weili Guan}, \bibinfo{person}{Yuan-Fang Li}, {and} \bibinfo{person}{Liqiang Nie}.} \bibinfo{year}{2024}\natexlab{b}.
\newblock \bibinfo{title}{MMGRec: Multimodal Generative Recommendation with Transformer Model}.
\newblock
\newblock
\showeprint[arxiv]{2404.16555}~[cs.IR]
\urldef\tempurl%
\url{https://arxiv.org/abs/2404.16555}
\showURL{%
\tempurl}


\bibitem[Liu et~al\mbox{.}(2024a)]%
        {multimodal-suvery}
\bibfield{author}{\bibinfo{person}{Qidong Liu}, \bibinfo{person}{Jiaxi Hu}, \bibinfo{person}{Yutian Xiao}, \bibinfo{person}{Xiangyu Zhao}, \bibinfo{person}{Jingtong Gao}, \bibinfo{person}{Wanyu Wang}, \bibinfo{person}{Qing Li}, {and} \bibinfo{person}{Jiliang Tang}.} \bibinfo{year}{2024}\natexlab{a}.
\newblock \showarticletitle{Multimodal recommender systems: A survey}.
\newblock \bibinfo{journal}{\emph{Comput. Surveys}} \bibinfo{volume}{57}, \bibinfo{number}{2} (\bibinfo{year}{2024}), \bibinfo{pages}{1--17}.
\newblock


\bibitem[Liu et~al\mbox{.}(2025)]%
        {modal_suvery_1}
\bibfield{author}{\bibinfo{person}{Qidong Liu}, \bibinfo{person}{Jiaxi Hu}, \bibinfo{person}{Yutian Xiao}, \bibinfo{person}{Xiangyu Zhao}, \bibinfo{person}{Jingtong Gao}, \bibinfo{person}{Wanyu Wang}, \bibinfo{person}{Qing Li}, {and} \bibinfo{person}{Jiliang Tang}.} \bibinfo{year}{2025}\natexlab{}.
\newblock \showarticletitle{Multimodal Recommender Systems: {A} Survey}.
\newblock \bibinfo{journal}{\emph{{ACM} Comput. Surv.}} \bibinfo{volume}{57}, \bibinfo{number}{2} (\bibinfo{year}{2025}), \bibinfo{pages}{26:1--26:17}.
\newblock


\bibitem[Liu et~al\mbox{.}(2018)]%
        {stamp}
\bibfield{author}{\bibinfo{person}{Qiao Liu}, \bibinfo{person}{Yifu Zeng}, \bibinfo{person}{Refuoe Mokhosi}, {and} \bibinfo{person}{Haibin Zhang}.} \bibinfo{year}{2018}\natexlab{}.
\newblock \showarticletitle{STAMP: Short-Term Attention/Memory Priority Model for Session-Based Recommendation}. In \bibinfo{booktitle}{\emph{KDD}}. \bibinfo{pages}{1831–1839}.
\newblock


\bibitem[Liu et~al\mbox{.}(2024c)]%
        {modal_suvery_2}
\bibfield{author}{\bibinfo{person}{Qijiong Liu}, \bibinfo{person}{Jieming Zhu}, \bibinfo{person}{Yanting Yang}, \bibinfo{person}{Quanyu Dai}, \bibinfo{person}{Zhaocheng Du}, \bibinfo{person}{Xiao{-}Ming Wu}, \bibinfo{person}{Zhou Zhao}, \bibinfo{person}{Rui Zhang}, {and} \bibinfo{person}{Zhenhua Dong}.} \bibinfo{year}{2024}\natexlab{c}.
\newblock \showarticletitle{Multimodal Pretraining, Adaptation, and Generation for Recommendation: {A} Survey}. In \bibinfo{booktitle}{\emph{KDD}}. \bibinfo{publisher}{{ACM}}, \bibinfo{pages}{6566--6576}.
\newblock


\bibitem[McAuley et~al\mbox{.}(2015)]%
        {amazon_1}
\bibfield{author}{\bibinfo{person}{Julian~J. McAuley}, \bibinfo{person}{Christopher Targett}, \bibinfo{person}{Qinfeng Shi}, {and} \bibinfo{person}{Anton van~den Hengel}.} \bibinfo{year}{2015}\natexlab{}.
\newblock \showarticletitle{Image-Based Recommendations on Styles and Substitutes}. In \bibinfo{booktitle}{\emph{SIGIR}}. \bibinfo{pages}{43--52}.
\newblock


\bibitem[Ni et~al\mbox{.}(2022)]%
        {sentence-t5}
\bibfield{author}{\bibinfo{person}{Jianmo Ni}, \bibinfo{person}{Gustavo~Hern{\'{a}}ndez {\'{A}}brego}, \bibinfo{person}{Noah Constant}, \bibinfo{person}{Ji Ma}, \bibinfo{person}{Keith~B. Hall}, \bibinfo{person}{Daniel Cer}, {and} \bibinfo{person}{Yinfei Yang}.} \bibinfo{year}{2022}\natexlab{}.
\newblock \showarticletitle{Sentence-T5: Scalable Sentence Encoders from Pre-trained Text-to-Text Models}. In \bibinfo{booktitle}{\emph{Findings of ACL}}. \bibinfo{publisher}{Association for Computational Linguistics}, \bibinfo{pages}{1864--1874}.
\newblock


\bibitem[Rajput et~al\mbox{.}(2023)]%
        {tiger}
\bibfield{author}{\bibinfo{person}{Shashank Rajput}, \bibinfo{person}{Nikhil Mehta}, \bibinfo{person}{Anima Singh}, \bibinfo{person}{Raghunandan~Hulikal Keshavan}, \bibinfo{person}{Trung Vu}, \bibinfo{person}{Lukasz Heldt}, \bibinfo{person}{Lichan Hong}, \bibinfo{person}{Yi Tay}, \bibinfo{person}{Vinh~Q. Tran}, \bibinfo{person}{Jonah Samost}, \bibinfo{person}{Maciej Kula}, \bibinfo{person}{Ed~H. Chi}, {and} \bibinfo{person}{Mahesh Sathiamoorthy}.} \bibinfo{year}{2023}\natexlab{}.
\newblock \showarticletitle{Recommender Systems with Generative Retrieval}. In \bibinfo{booktitle}{\emph{NeurIPS}}.
\newblock


\bibitem[Rendle et~al\mbox{.}(2010)]%
        {fpmc}
\bibfield{author}{\bibinfo{person}{Steffen Rendle}, \bibinfo{person}{Christoph Freudenthaler}, {and} \bibinfo{person}{Lars Schmidt-Thieme}.} \bibinfo{year}{2010}\natexlab{}.
\newblock \showarticletitle{Factorizing Personalized Markov Chains for Next-basket Recommendation}. In \bibinfo{booktitle}{\emph{WWW}}. \bibinfo{pages}{811--820}.
\newblock


\bibitem[Shang et~al\mbox{.}(2023)]%
        {micro_video_1}
\bibfield{author}{\bibinfo{person}{Yu Shang}, \bibinfo{person}{Chen Gao}, \bibinfo{person}{Jiansheng Chen}, \bibinfo{person}{Depeng Jin}, \bibinfo{person}{Meng Wang}, {and} \bibinfo{person}{Yong Li}.} \bibinfo{year}{2023}\natexlab{}.
\newblock In \bibinfo{booktitle}{\emph{SIGIR}}. \bibinfo{publisher}{{ACM}}, \bibinfo{pages}{433--442}.
\newblock


\bibitem[Song et~al\mbox{.}(2023)]%
        {fashion_rec_2}
\bibfield{author}{\bibinfo{person}{Xuemeng Song}, \bibinfo{person}{Chun Wang}, \bibinfo{person}{Changchang Sun}, \bibinfo{person}{Shanshan Feng}, \bibinfo{person}{Min Zhou}, {and} \bibinfo{person}{Liqiang Nie}.} \bibinfo{year}{2023}\natexlab{}.
\newblock \showarticletitle{MM-FRec: Multi-Modal Enhanced Fashion Item Recommendation}.
\newblock \bibinfo{journal}{\emph{{IEEE} Trans. Knowl. Data Eng.}} \bibinfo{volume}{35}, \bibinfo{number}{10} (\bibinfo{year}{2023}), \bibinfo{pages}{10072--10084}.
\newblock


\bibitem[Sun et~al\mbox{.}(2019)]%
        {bert4rec}
\bibfield{author}{\bibinfo{person}{Fei Sun}, \bibinfo{person}{Jun Liu}, \bibinfo{person}{Jian Wu}, \bibinfo{person}{Changhua Pei}, \bibinfo{person}{Xiao Lin}, \bibinfo{person}{Wenwu Ou}, {and} \bibinfo{person}{Peng Jiang}.} \bibinfo{year}{2019}\natexlab{}.
\newblock \showarticletitle{BERT4Rec: Sequential Recommendation with Bidirectional Encoder Representations from Transformer}. In \bibinfo{booktitle}{\emph{CIKM}}. \bibinfo{pages}{1441--1450}.
\newblock


\bibitem[van~den Oord et~al\mbox{.}(2018)]%
        {infonce}
\bibfield{author}{\bibinfo{person}{A{\"{a}}ron van~den Oord}, \bibinfo{person}{Yazhe Li}, {and} \bibinfo{person}{Oriol Vinyals}.} \bibinfo{year}{2018}\natexlab{}.
\newblock \showarticletitle{Representation Learning with Contrastive Predictive Coding}.
\newblock \bibinfo{journal}{\emph{CoRR}}  \bibinfo{volume}{abs/1807.03748} (\bibinfo{year}{2018}).
\newblock
\showeprint[arXiv]{1807.03748}
\urldef\tempurl%
\url{http://arxiv.org/abs/1807.03748}
\showURL{%
\tempurl}


\bibitem[van~den Oord et~al\mbox{.}(2017)]%
        {vq}
\bibfield{author}{\bibinfo{person}{A{\"{a}}ron van~den Oord}, \bibinfo{person}{Oriol Vinyals}, {and} \bibinfo{person}{Koray Kavukcuoglu}.} \bibinfo{year}{2017}\natexlab{}.
\newblock \showarticletitle{Neural Discrete Representation Learning}. In \bibinfo{booktitle}{\emph{NeurIPS}}. \bibinfo{pages}{6306--6315}.
\newblock


\bibitem[Wang et~al\mbox{.}(2023)]%
        {missrec}
\bibfield{author}{\bibinfo{person}{Jinpeng Wang}, \bibinfo{person}{Ziyun Zeng}, \bibinfo{person}{Yunxiao Wang}, \bibinfo{person}{Yuting Wang}, \bibinfo{person}{Xingyu Lu}, \bibinfo{person}{Tianxiang Li}, \bibinfo{person}{Jun Yuan}, \bibinfo{person}{Rui Zhang}, \bibinfo{person}{Hai{-}Tao Zheng}, {and} \bibinfo{person}{Shu{-}Tao Xia}.} \bibinfo{year}{2023}\natexlab{}.
\newblock \showarticletitle{MISSRec: Pre-training and Transferring Multi-modal Interest-aware Sequence Representation for Recommendation}. In \bibinfo{booktitle}{\emph{MM}}. \bibinfo{publisher}{{ACM}}, \bibinfo{pages}{6548--6557}.
\newblock


\bibitem[Wang et~al\mbox{.}(2019)]%
        {ugrec}
\bibfield{author}{\bibinfo{person}{Pengfei Wang}, \bibinfo{person}{Hanxiong Chen}, \bibinfo{person}{Yadong Zhu}, \bibinfo{person}{Huawei Shen}, {and} \bibinfo{person}{Yongfeng Zhang}.} \bibinfo{year}{2019}\natexlab{}.
\newblock \showarticletitle{Unified Collaborative Filtering over Graph Embeddings}. In \bibinfo{booktitle}{\emph{SIGIR}}. \bibinfo{pages}{155--164}.
\newblock


\bibitem[Wang et~al\mbox{.}(2015)]%
        {hrm}
\bibfield{author}{\bibinfo{person}{Pengfei Wang}, \bibinfo{person}{Jiafeng Guo}, \bibinfo{person}{Yanyan Lan}, \bibinfo{person}{Jun Xu}, \bibinfo{person}{Shengxian Wan}, {and} \bibinfo{person}{Xueqi Cheng}.} \bibinfo{year}{2015}\natexlab{}.
\newblock \showarticletitle{Learning Hierarchical Representation Model for NextBasket Recommendation}. In \bibinfo{booktitle}{\emph{SIGIR}}. \bibinfo{pages}{403--412}.
\newblock


\bibitem[Wang et~al\mbox{.}(2024b)]%
        {asif}
\bibfield{author}{\bibinfo{person}{Shuhan Wang}, \bibinfo{person}{Bin Shen}, \bibinfo{person}{Xu Min}, \bibinfo{person}{Yong He}, \bibinfo{person}{Xiaolu Zhang}, \bibinfo{person}{Liang Zhang}, \bibinfo{person}{Jun Zhou}, {and} \bibinfo{person}{Linjian Mo}.} \bibinfo{year}{2024}\natexlab{b}.
\newblock \showarticletitle{Aligned side information fusion method for sequential recommendation}. In \bibinfo{booktitle}{\emph{Companion Proceedings of the ACM Web Conference 2024}}. \bibinfo{pages}{112--120}.
\newblock


\bibitem[Wang et~al\mbox{.}(2024a)]%
        {letter}
\bibfield{author}{\bibinfo{person}{Wenjie Wang}, \bibinfo{person}{Honghui Bao}, \bibinfo{person}{Xinyu Lin}, \bibinfo{person}{Jizhi Zhang}, \bibinfo{person}{Yongqi Li}, \bibinfo{person}{Fuli Feng}, \bibinfo{person}{See{-}Kiong Ng}, {and} \bibinfo{person}{Tat{-}Seng Chua}.} \bibinfo{year}{2024}\natexlab{a}.
\newblock \showarticletitle{Learnable Item Tokenization for Generative Recommendation}. In \bibinfo{booktitle}{\emph{CIKM}}. \bibinfo{publisher}{{ACM}}, \bibinfo{pages}{2400--2409}.
\newblock


\bibitem[Wang et~al\mbox{.}(2024c)]%
        {eager}
\bibfield{author}{\bibinfo{person}{Ye Wang}, \bibinfo{person}{Jiahao Xun}, \bibinfo{person}{Minjie Hong}, \bibinfo{person}{Jieming Zhu}, \bibinfo{person}{Tao Jin}, \bibinfo{person}{Wang Lin}, \bibinfo{person}{Haoyuan Li}, \bibinfo{person}{Linjun Li}, \bibinfo{person}{Yan Xia}, \bibinfo{person}{Zhou Zhao}, {and} \bibinfo{person}{Zhenhua Dong}.} \bibinfo{year}{2024}\natexlab{c}.
\newblock \showarticletitle{{EAGER:} Two-Stream Generative Recommender with Behavior-Semantic Collaboration}. In \bibinfo{booktitle}{\emph{KDD}}. \bibinfo{publisher}{{ACM}}, \bibinfo{pages}{3245--3254}.
\newblock


\bibitem[Wei et~al\mbox{.}(2021)]%
        {grcn}
\bibfield{author}{\bibinfo{person}{Yinwei Wei}, \bibinfo{person}{Xiang Wang}, \bibinfo{person}{Liqiang Nie}, \bibinfo{person}{Xiangnan He}, {and} \bibinfo{person}{Tat{-}Seng Chua}.} \bibinfo{year}{2021}\natexlab{}.
\newblock \showarticletitle{{GRCN:} Graph-Refined Convolutional Network for Multimedia Recommendation with Implicit Feedback}.
\newblock \bibinfo{journal}{\emph{CoRR}}  \bibinfo{volume}{abs/2111.02036} (\bibinfo{year}{2021}).
\newblock
\showeprint[arXiv]{2111.02036}
\urldef\tempurl%
\url{https://arxiv.org/abs/2111.02036}
\showURL{%
\tempurl}


\bibitem[Wei et~al\mbox{.}(2019)]%
        {mmgcn}
\bibfield{author}{\bibinfo{person}{Yinwei Wei}, \bibinfo{person}{Xiang Wang}, \bibinfo{person}{Liqiang Nie}, \bibinfo{person}{Xiangnan He}, \bibinfo{person}{Richang Hong}, {and} \bibinfo{person}{Tat{-}Seng Chua}.} \bibinfo{year}{2019}\natexlab{}.
\newblock \showarticletitle{{MMGCN:} Multi-modal Graph Convolution Network for Personalized Recommendation of Micro-video}. In \bibinfo{booktitle}{\emph{MM}}. \bibinfo{pages}{1437--1445}.
\newblock


\bibitem[Xia et~al\mbox{.}(2023)]%
        {cross_modal}
\bibfield{author}{\bibinfo{person}{Yan Xia}, \bibinfo{person}{Hai Huang}, \bibinfo{person}{Jieming Zhu}, {and} \bibinfo{person}{Zhou Zhao}.} \bibinfo{year}{2023}\natexlab{}.
\newblock \showarticletitle{Achieving Cross Modal Generalization with Multimodal Unified Representation}. In \bibinfo{booktitle}{\emph{NeurIPS}}.
\newblock


\bibitem[Xie et~al\mbox{.}(2022)]%
        {difsr}
\bibfield{author}{\bibinfo{person}{Yueqi Xie}, \bibinfo{person}{Peilin Zhou}, {and} \bibinfo{person}{Sunghun Kim}.} \bibinfo{year}{2022}\natexlab{}.
\newblock In \bibinfo{booktitle}{\emph{SIGIR}}. \bibinfo{publisher}{{ACM}}, \bibinfo{pages}{1611--1621}.
\newblock


\bibitem[Xie et~al\mbox{.}(2021)]%
        {dev_1}
\bibfield{author}{\bibinfo{person}{Zhe Xie}, \bibinfo{person}{Chengxuan Liu}, \bibinfo{person}{Yichi Zhang}, \bibinfo{person}{Hongtao Lu}, \bibinfo{person}{Dong Wang}, {and} \bibinfo{person}{Yue Ding}.} \bibinfo{year}{2021}\natexlab{}.
\newblock \showarticletitle{Adversarial and Contrastive Variational Autoencoder for Sequential Recommendation}. In \bibinfo{booktitle}{\emph{WWW}}. \bibinfo{pages}{449--459}.
\newblock


\bibitem[Xv et~al\mbox{.}(2024)]%
        {xv2024improving}
\bibfield{author}{\bibinfo{person}{Guipeng Xv}, \bibinfo{person}{Xinyu Li}, \bibinfo{person}{Ruobing Xie}, \bibinfo{person}{Chen Lin}, \bibinfo{person}{Chong Liu}, \bibinfo{person}{Feng Xia}, \bibinfo{person}{Zhanhui Kang}, {and} \bibinfo{person}{Leyu Lin}.} \bibinfo{year}{2024}\natexlab{}.
\newblock \showarticletitle{Improving Multi-modal Recommender Systems by Denoising and Aligning Multi-modal Content and User Feedback}. In \bibinfo{booktitle}{\emph{Proceedings of the 30th ACM SIGKDD Conference on Knowledge Discovery and Data Mining}}. \bibinfo{pages}{3645--3656}.
\newblock


\bibitem[Yi et~al\mbox{.}(2022a)]%
        {micro_video_2}
\bibfield{author}{\bibinfo{person}{Zixuan Yi}, \bibinfo{person}{Xi Wang}, \bibinfo{person}{Iadh Ounis}, {and} \bibinfo{person}{Craig MacDonald}.} \bibinfo{year}{2022}\natexlab{a}.
\newblock In \bibinfo{booktitle}{\emph{SIGIR}}. \bibinfo{publisher}{{ACM}}, \bibinfo{pages}{1807--1811}.
\newblock


\bibitem[Yi et~al\mbox{.}(2022b)]%
        {mmgcl}
\bibfield{author}{\bibinfo{person}{Zixuan Yi}, \bibinfo{person}{Xi Wang}, \bibinfo{person}{Iadh Ounis}, {and} \bibinfo{person}{Craig MacDonald}.} \bibinfo{year}{2022}\natexlab{b}.
\newblock \showarticletitle{Multi-modal Graph Contrastive Learning for Micro-video Recommendation}. In \bibinfo{booktitle}{\emph{SIGIR}}. \bibinfo{publisher}{{ACM}}, \bibinfo{pages}{1807--1811}.
\newblock


\bibitem[Yu et~al\mbox{.}(2023)]%
        {mvgcn}
\bibfield{author}{\bibinfo{person}{Penghang Yu}, \bibinfo{person}{Zhiyi Tan}, \bibinfo{person}{Guanming Lu}, {and} \bibinfo{person}{Bing-Kun Bao}.} \bibinfo{year}{2023}\natexlab{}.
\newblock \showarticletitle{Multi-View Graph Convolutional Network for Multimedia Recommendation}. In \bibinfo{booktitle}{\emph{Proceedings of the 31st ACM International Conference on Multimedia}} \emph{(\bibinfo{series}{MM '23})}. \bibinfo{pages}{6576–6585}.
\newblock


\bibitem[Yuan et~al\mbox{.}(2023)]%
        {id_vs_modal}
\bibfield{author}{\bibinfo{person}{Zheng Yuan}, \bibinfo{person}{Fajie Yuan}, \bibinfo{person}{Yu Song}, \bibinfo{person}{Youhua Li}, \bibinfo{person}{Junchen Fu}, \bibinfo{person}{Fei Yang}, \bibinfo{person}{Yunzhu Pan}, {and} \bibinfo{person}{Yongxin Ni}.} \bibinfo{year}{2023}\natexlab{}.
\newblock \showarticletitle{Where to Go Next for Recommender Systems? {ID-} vs. Modality-based Recommender Models Revisited}. In \bibinfo{booktitle}{\emph{SIGIR}}. \bibinfo{publisher}{{ACM}}, \bibinfo{pages}{2639--2649}.
\newblock


\bibitem[Zeghidour et~al\mbox{.}(2022)]%
        {rqvae}
\bibfield{author}{\bibinfo{person}{Neil Zeghidour}, \bibinfo{person}{Alejandro Luebs}, \bibinfo{person}{Ahmed Omran}, \bibinfo{person}{Jan Skoglund}, {and} \bibinfo{person}{Marco Tagliasacchi}.} \bibinfo{year}{2022}\natexlab{}.
\newblock \showarticletitle{SoundStream: An End-to-End Neural Audio Codec}.
\newblock \bibinfo{journal}{\emph{{IEEE} {ACM} Trans. Audio Speech Lang. Process.}}  \bibinfo{volume}{30} (\bibinfo{year}{2022}), \bibinfo{pages}{495--507}.
\newblock


\bibitem[Zhang et~al\mbox{.}(2019)]%
        {fdsa}
\bibfield{author}{\bibinfo{person}{Tingting Zhang}, \bibinfo{person}{Pengpeng Zhao}, \bibinfo{person}{Yanchi Liu}, \bibinfo{person}{Victor~S. Sheng}, \bibinfo{person}{Jiajie Xu}, \bibinfo{person}{Deqing Wang}, \bibinfo{person}{Guanfeng Liu}, {and} \bibinfo{person}{Xiaofang Zhou}.} \bibinfo{year}{2019}\natexlab{}.
\newblock \showarticletitle{Feature-level Deeper Self-Attention Network for Sequential Recommendation}. In \bibinfo{booktitle}{\emph{IJCAI}}. \bibinfo{pages}{4320--4326}.
\newblock


\bibitem[Zhao et~al\mbox{.}(2021)]%
        {recbole}
\bibfield{author}{\bibinfo{person}{Wayne~Xin Zhao}, \bibinfo{person}{Shanlei Mu}, \bibinfo{person}{Yupeng Hou}, \bibinfo{person}{Zihan Lin}, \bibinfo{person}{Yushuo Chen}, \bibinfo{person}{Xingyu Pan}, \bibinfo{person}{Kaiyuan Li}, \bibinfo{person}{Yujie Lu}, \bibinfo{person}{Hui Wang}, \bibinfo{person}{Changxin Tian}, \bibinfo{person}{Yingqian Min}, \bibinfo{person}{Zhichao Feng}, \bibinfo{person}{Xinyan Fan}, \bibinfo{person}{Xu Chen}, \bibinfo{person}{Pengfei Wang}, \bibinfo{person}{Wendi Ji}, \bibinfo{person}{Yaliang Li}, \bibinfo{person}{Xiaoling Wang}, {and} \bibinfo{person}{Ji{-}Rong Wen}.} \bibinfo{year}{2021}\natexlab{}.
\newblock \showarticletitle{RecBole: Towards a Unified, Comprehensive and Efficient Framework for Recommendation Algorithms}. In \bibinfo{booktitle}{\emph{CIKM}}. \bibinfo{pages}{4653--4664}.
\newblock


\bibitem[Zhou et~al\mbox{.}(2020)]%
        {s3rec}
\bibfield{author}{\bibinfo{person}{Kun Zhou}, \bibinfo{person}{Hui Wang}, \bibinfo{person}{Wayne~Xin Zhao}, \bibinfo{person}{Yutao Zhu}, \bibinfo{person}{Sirui Wang}, \bibinfo{person}{Fuzheng Zhang}, \bibinfo{person}{Zhongyuan Wang}, {and} \bibinfo{person}{Ji{-}Rong Wen}.} \bibinfo{year}{2020}\natexlab{}.
\newblock \showarticletitle{S3-Rec: Self-Supervised Learning for Sequential Recommendation with Mutual Information Maximization}. In \bibinfo{booktitle}{\emph{CIKM}}. \bibinfo{pages}{1893--1902}.
\newblock


\bibitem[Zhu et~al\mbox{.}(2024)]%
        {cost}
\bibfield{author}{\bibinfo{person}{Jieming Zhu}, \bibinfo{person}{Mengqun Jin}, \bibinfo{person}{Qijiong Liu}, \bibinfo{person}{Zexuan Qiu}, \bibinfo{person}{Zhenhua Dong}, {and} \bibinfo{person}{Xiu Li}.} \bibinfo{year}{2024}\natexlab{}.
\newblock \showarticletitle{CoST: Contrastive Quantization based Semantic Tokenization for Generative Recommendation}. In \bibinfo{booktitle}{\emph{RecSys}}. \bibinfo{publisher}{{ACM}}, \bibinfo{pages}{969--974}.
\newblock


\end{thebibliography}

\end{document}